\documentclass{article}%<<<1
\usepackage{unicode}
\DeclareUnicodeCharacter{22F1}{\smash\ddots}
\DeclareUnicodeCharacter{22F0}{\smash\iddots}
\DeclareUnicodeCharacter{2264}{\leqslant}
\DeclareUnicodeCharacter{2265}{\geqslant}
\usepackage{amsmath,amssymb,mathrsfs}
\usepackage{bm}
\usepackage{lmodern}
\usepackage{mathdots}
\usepackage{algorithm}
\usepackage{algpseudocode}
\usepackage[all]{xy}
\usepackage{graphicx}

\usepackage{amsthm}
\usepackage[margin=25mm]{geometry}

\usepackage{color}
\newif\ifapx \apxtrue % If appendix

\def\linkcounter#1#2{\edef\magic{\noexpand\let
  \expandafter\noexpand\csname c@#1\endcsname
  \expandafter\noexpand\csname c@#2\endcsname}\magic}
\def\newthm#1#2{\newtheorem{#1}{#2}[section]\linkcounter{#1}{equation}}
\newthm{prop}{Proposition}
\newthm{thm}{Theorem}
\newthm{lem}{Lemma}

\def\labelenumi{(\roman{enumi})}
\def\itemref#1{\expandafter\ifx\csname r@#1\endcsname\relax
  {\bfseries ??}\else{\setcounter{enumi}{\ref{#1}}\labelenumi}\fi}
\let\fr\mathfrak

\def\transpose#1{{\vphantom{#1}}^{\mathrm{t}}\!#1}
\def\pa#1{\left(#1\right)}
\def\floor#1{\left\lfloor#1\right\rfloor}
\def\chev#1{\left\langle#1\right\rangle}
\def\acco#1{\left\{#1\right\}}
\def\abs#1{\left|#1\right|}
\def\mat#1{\begin{pmatrix}#1\end{pmatrix}}
\def\smat{\def\arraystretch{.66}\mat}
\def\card#1{\abs{#1}}
\def\chk#1{#1^{\smash{\scalebox{.7}[1.4]{\rotatebox{90}{\guilsinglleft}}}}}
\DeclareMathOperator\Ker{Ker}
\DeclareMathOperator\Hom{Hom}

\DeclareMathOperator\GL{GL}
\DeclareMathOperator\PGL{PGL}
\DeclareMathOperator\Tr{Tr}
\def\F{\mathbb{F}}

\def\Ot{\widetilde{O}}

\def\gitkw$#1:#2${{\small \textbf{#1: }\texttt{#2}}}
\makeatletter
\def\clap #1{\hbox to 0pt{\hss#1\hss}}
\def\stretchdots#1#2#3#4{
  \setbox0=\hbox{$#4$}\dimen0= \wd0 \advance \dimen0 2\arraycolsep
  \rlap{\kern -\arraycolsep \hbox to \dimen0 {%
  \hss \raise #1 \clap{$.$}\hss
  \hss \raise #2 \clap{$\m@th.$}\hss
  \hss \raise #3 \clap{$\m@th.$}\hss}}%
  \kern \wd0
}
\def\siddots{\stretchdots{0pt}{4pt}{8pt}}
\def\sddots{\stretchdots{8pt}{4pt}{0pt}}
\makeatother%>>>

\begin{document}
\title%[Solving ``Isomorphism of Polynomials with Two Secrets'']%
{Solving the ``Isomorphism of Polynomials with Two Secrets'' Problem
for All Pairs of Quadratic Forms}%<<<1
\author{Jérôme Plût\footnote{ANSSI}
\and Pierre-Alain Fouque\footnote{Université Rennes 1 and Institut
Universitaire de France}
\and Gilles Macario-Rat\footnote{Orange Labs}}
\maketitle

\centerline{\gitkw $Id:0ba9f0d$
\gitkw $Author:jerome.plut@ssi.gouv.fr$
\gitkw $Date:2014-dec-11 08:46:48 +0100$}

\begin{abstract}%<<<
We study the Isomorphism of Polynomial (IP2S) problem
with~$m=2$ homogeneous quadratic polynomials of $n$ variables over a finite field of odd
characteristic: given two quadratic polynomials $(\bm{a},\bm{b})$ 
on $n$ variables, we find two bijective linear maps $(s,t)$ such that
$\bm{b}=t\circ \bm{a}\circ s$. We give an algorithm computing~$s$ and~$t$
in time complexity~$\Ot(n^4)$ for all instances.

The IP2S problem was introduced in cryptography by Patarin back in 1996.
The special case of this problem when $t$ is the identity is called
the isomorphism with one secret (IP1S) problem.
Generic algebraic equation solvers (for example using Gröbner bases)
solve quite well random instances of the IP1S problem. For the particular
\emph{cyclic} instances of IP1S, a cubic-time algorithm was later
given~\cite{MPG2013} and explained in terms of pencils of quadratic forms
over all finite fields; in particular, the cyclic IP1S problem in odd
characteristic reduces to the computation of the square root of a matrix.

We give here an algorithm solving all cases of the IP1S problem in odd
characteristic using two new tools, the Kronecker form for a singular
quadratic pencil, and the reduction of bilinear forms over a non-commutative
algebra. Finally, we show that the second secret in the IP2S problem may
be recovered in cubic time.
\end{abstract}%>>>
\section*{Introduction}%<<<1
\subsection*{The IP1S and IP2S problems}%<<<2

The \emph{Isomorphism of Polynomial with Two Secrets} (IP2S) problem is
the following: given a field~$k$ and two $m$-uples $\bm{a} = (a_1, …,
a_m)$ and~$\bm{b} = (b_1, …, b_m)$ in $n$~variables~$(x_1, …, x_n)$, compute
two invertible linear maps~$s ∈ \GL_n(k)$ of the variables~$x_i$
and~$t ∈ \GL_m(k)$ of the polynomials~$a_i$ such that
\begin{equation*}
\bm{b} = t ∘ \bm{a} ∘ s.
\end{equation*}
The particular case where we restrict~$t$ to the identity transformation
is also known as the \emph{Isomorphism of Polynomials with One Secret}
(IP1S). Both these problems have been introduced in cryptography by
Patarin in~\cite{DBLP:conf/eurocrypt/Patarin96} to construct an efficient
authentication scheme, 
as an alternative to the Graph Isomorphism Problem (GI) proposed by Goldreich, Micali and Wigderson~\cite{DBLP:journals/jacm/GoldreichMW91}.
The IP problem was appealing since 
it seems more difficult than the Graph Isomorphism problem~\cite{DBLP:conf/eurocrypt/PatarinGC98}. 
Agrawal and Saxena reduced~\cite{DBLP:conf/stacs/AgrawalS06} the Graph
Isomorphism problem to the particular case of IP1S using two polynomials,
one of them being a quadratic form encoding the adjacency matrix of the
graph, and the other one being the cubic~$\sum x_i^3$, over a finite field of
odd characteristic.
For the case of quadratic polynomials, the status of this problem is unclear despite recent 
intensive research in the cryptographic community since this case is the most interesting 
for practical schemes. There exists a claimed reduction between the quadratic IP1S problem and the GI
problem~\cite{DBLP:conf/eurocrypt/PatarinGC98}, but we realized that this proof is incomplete. 
Indeed, the proof works by induction and decomposes any permutation as the composition of 
transpositions. It is possible to write a system of quadratic polynomials such that the only solutions 
of the IP1S problem will be the identity or a transposition by modifying a bit the systems proposed 
in~\cite{DBLP:conf/eurocrypt/PatarinGC98}. However, it is not obvious how we can compose the systems 
of equations such that the solutions will be the composition of the solutions. 

\bigbreak

The defining parameters of the IP problems are the number~$n$ of
variables, the number~$m$ of polynomials, their degree, and the finite
field~$k$. For
efficiency reasons, the degree is generally small, involving only
quadratic and cubic equations. To our knowledge, no significant progress
has been done on the cubic case.

We limit ourselves to the special case of two equations, both of which
being homogeneous polynomials of degree~two. According to previous
literature~\cite{DBLP:conf/eurocrypt/Perret05,DBLP:conf/eurocrypt/FaugereP06,DBLP:conf/pkc/BouillaguetFFP11,DBLP:conf/eurocrypt/BouillaguetFV13},
this is the most difficult case.

The case with only one homogeneous quadratic equation amounts to 
reduction of quadratic forms, which has been known for
centuries~\cite{gauss,lidl1997finite}. In the non-homogeneous case, the
presence of affine terms gives linear relations between the secret
unknowns~\cite{DBLP:conf/eurocrypt/PatarinGC98}, and this extra
information actually helps generic solvers, for example those using
Gröbner bases~\cite{DBLP:conf/eurocrypt/FaugereP06}.
The case with more than two equations is easier since we can relinearize
the systems~\cite{DBLP:conf/pkc/BouillaguetFFP11}.

\subsection*{Previous work}%<<<2

Some recent advances have been made on the IP1S problem in the case of
two
homogeneous quadratic equations.

Bouillaguet, Fouque and Macario-Rat
in 2011~\cite{DBLP:conf/asiacrypt/BouillaguetFM11} 
used \emph{pencils} of quadratic forms,
which are pairs~$(b_{∞}, b_0)$ of such forms,
to recover the secret mappings $s$ and $t$ when three equations are available
and one of the quadratic equations $\bm{a}$
comes from a special mapping $X\mapsto X^{q^\theta+1}$ over $\F_q$.
In the case of the IP problem,
this is optimal using an information theoretic argument.

A case of interest is the particular case of \emph{cyclic} pencils: a
pencil~$\bm{b} = (b_{∞}, b_0)$ is \emph{cyclic} if $b_{∞}$~is invertible
and $b_{∞}^{-1} b_{0}$ is a cyclic matrix, \emph{i.e.} its characteristic
polynomial is equal to its minimal polynomial. The cyclic case is
dominant (it is defined by the non-cancellation of some polynomial
functions of the coefficients of~$\bm{b}$). For cyclic instances of the
IP1S problem, the Gröbner basis approach works
well~\cite{DBLP:conf/pkc/BouillaguetFFP11} since the number of solutions
is known to be small. For all other instances, the number of solutions is
empirically large and such algorithms are then well known to be less
efficient.
Macario-Rat, Plût and Gilbert gave in 2013~\cite{MPG2013}
an algebraic solution to the cyclic instances of the IP1S problem
for $m=2$ over finite fields of any characteristic.
% Macario-Rat, Plût and Gilbert explained in 2013~\cite{MPG2013} how to
% solve cyclic instances of the IP1S problem for~$m=2$ over finite fields
% of any characteristic. A pencil~$\bm{b} = (b_{∞}, b_0)$ is \emph{cyclic}
% if $b_{∞}$~is invertible and~$b_{∞}^{-1} b_{0}$ is a cyclic matrix,
% \emph{i.e.} its characteristic polynomial is equal to its minimal
% polynomial. Although the cyclic case is dominant (it is defined by the
% non-cancellation of some polynomial functions
% of the coefficients of~$\bm{b}$), it is not the general case in a
% practical sense. In the cyclic instances, Gröbner bases works
% well~\cite{DBLP:conf/pkc/BouillaguetFFP11} since in this case, the number
% of solutions is small. For all other instances, the number of solutions
% is large, and it is well-known that in this case, such algorithms are
% less efficient.

Finally, Berthomieu, Faugère and Perret
proposed in 2014~\cite{DBLP:journals/corr/BerthomieuFP13}
a polynomial algorithm for IP1S with any number of equations when~$2 ≠ 0$.
Given two families of polynomials over a field~$k$,
they give a solution to the IP1S problem
over a tower~$k'$ of real quadratic extensions of~$k$
(a \emph{real quadratic extension} being obtained by
adjoining the square root of a sum of squares).
This solves the IP1S problem over the original field~$k$ only if
$k$~is \emph{Euclidean}, \emph{i.e.} has no real quadratic extension.
This is the case for example if $k$~is a closed real field such as~$ℝ$
or the field~$ℝ_{\mathrm{alg}}$ of real algebraic numbers,
or an algebraically closed field;
since any quadratic extension of a finite field is real,
no finite field is Euclidean.

% Over the
% original field, this solves the \emph{decisional} IP1S problem:
% determining whether two families of polynomials are isomorphic. This
% latter problem does not have many cryptographic applications.

\subsection*{Our contributions}%<<<2
This work covers the IP1S and IP2S problems
for~$m = 2$ homogeneous quadratic equations (\emph{quadratic pencils})
over a finite field of any characteristic.

For the IP1S problem in odd characteristic, we introduce three new tools.
A quadratic pencil is \emph{regular}
if its characteristic polynomial is not zero, and \emph{singular} otherwise.
We first give a full description of the \emph{singular part}
of all quadratic pencils in section~\ref{S:bil-sing}.
This is the same as the classical Kronecker classification of quadratic
pencils.
Although this dates back to Kronecker, the original proof used real and
complex analysis (square roots of matrices);
our new proof is algorithmic and valid over any field.

For regular pencils, we then give a decomposition as the orthogonal
direct sum of \emph{local pencils}, for which we know that at least one
of the quadratic forms is regular.

We finally prove that the IP1S problem
for a local, regular pencil over a non-binary field
amounts to a reduction problem for some quadratic forms over a local algebra.
As this is a well-known theory (in odd characteristic), we are able to
give a polynomial-time answer to all instances of IP1S in
section~\ref{S:bil-regular}. Our proof here specializes to that
of~\cite{MPG2013} in the particular case of a cyclic pencil.

The characteristic-two case is different from the odd-characteristic case
in that quadratic forms are no longer determined by their polar forms.
Although we were unable to give a full classification
for quadratic pencils in characteristic two,
we use the description of automorphisms of the polar pencil
to give a polynomial-time algorithm computing isomorphisms
between the quadratic pencils in section~\ref{S:quad-reg}.

\medskip

The last section explains how we recover the second (``outer'') secret
in the two-secret IP2S problem.
Since applying an outer linear combination to a pencil
leaves the singular part of the pencil unchanged (up to isomorphism),
we can use the regular part alone to recover the outer secret.
This is done using the factorization of the characteristic polynomial.

\subsection*{Mathematical background and notations}%<<<2
Throughout this document, $k$~is a finite field.
Let~$V$ be a $n$-dimensional vector space over the field~$k$. We study
the IP1S and IP2S problems for \emph{quadratic forms} on~$V$, which are
homogeneous polynomials of degree~$2$ in some coordinates on~$V$. To a
quadratic form~$q$, one may associate the \emph{polar form}~$b$ defined
by
\begin{equation*}\label{eq:polar}
b(x,y) = q(x+y) - q(x) - q(y);
\end{equation*}
this is a symmetric bilinear forms, and it satisfies the \emph{polarity
identity}
\begin{equation*}\label{eq:polarity}
b(x,x) = 2q(x).
\end{equation*}
If~$2 ≠ 0$ in~$k$, then the polarity identity is a bijection between
quadratic forms and bilinear forms. Therefore, instead of quadratic
forms, we shall study bilinear forms.

In the case where~$2 = 0$ in~$k$, the situation is more complicated;
the polarity identity is no longer a bijection, but polar forms are
instead alternating bilinear forms. This means that their classification is
quite different from the odd-characteristic case~\cite{milnorhusemoller},
and relies on the Arf invariant.

Let~$\chk{V}$ be the dual of the vector space~$V$. A bilinear form~$b$
on~$V$ is the same as a linear map~$b: V → \chk{V}$. The bilinear form is
\emph{regular} if it defines an invertible linear map~$V → \chk{V}$. In
this case, for any endomorphism~$u$ of~$V$, there exists a unique
endomorphism~$u^{⋆}$ of~$V$ such that~$b(x,u(y)) = b(u^{⋆}(x), y)$; the
endomorphism~$u^{⋆}$ is called the \emph{left-adjoint} of~$u$. If $b$~is
symmetric then left- and right-adjoints coincide.

An \emph{(affine) pencil of symmetric bilinear forms} over~$V$, or a
\emph{symmetric pencil} in short, is a pair of symmetric bilinear
forms~$\bm{b} = (b_{∞}, b_{0})$ over~$V$. We write this pencil in affine
form as~$b_{λ} = λ b_{∞} + b_{0}$, and in projective form as~$b_{λ:μ} = λ
b_{∞} + μ b_{0}$, where $b_{0}$ and~$b_{∞}$ are symmetric bilinear forms.
Given two vector spaces equipped with pencils~$(V, b)$ and~$(V', b')$, a
\emph{linear map} of pencils is a linear map~$s: V → V'$ such
that~$b'_{λ} ∘ s = b_{λ}$. A \emph{projective map} of pencils is a
pair~$(s, t)$, where $s: V → V'$ is a linear map and $t ∈ \PGL_2 (k)$~is
a homography such that~$b'_{λ} ∘ s = b_{t(λ)}$. We define quadratic
pencils in the same way. The IP1S problem is then the computation of a
linear isomorphism of quadratic pencils, whereas the IP2S problem is the
computation of a projective isomorphism.

Two elements~$x$ and~$y$ of~$V$ are \emph{orthogonal} for a
bilinear form~$b$ if~$b(x,y) = 0$. They are orthogonal
for a pencil~$(b_{λ})$ if, for all~$λ$, $b_{λ}(x,y) = 0$. We write~$x
⟂_{b} y$, or $x ⟂ y$ when the bilinear form or pencil is clear from
context. We write~$W^{⟂}$ for the orthogonal of a subspace~$W ⊂ V$.
A space~$W$ is \emph{self-orthogonal} if~$W ⊂ W^{⟂}$.

\medbreak

We write~$R^{m×n}$ for the vector space of matrices with entries in~$R$
having $m$~lines and $n$~columns, and $\transpose{A}$~for the transpose
of a matrix~$A$. A \emph{symmetric} matrix is a matrix such that~$A =
\transpose{A}$. Symmetric bilinear forms~$b$ correspond to symmetric
matrices~$B$. A bilinear form is regular iff its matrix is invertible.
For any endomorphism~$u$ with matrix~$U$, the adjoint endomorphism
(relatively to~$B$) has matrix~$U^{⋆} = B^{-1} · \transpose{U} · B$.
The \emph{companion matrix}~$M_f$ of a polynomial~$f$
is the matrix of multiplication by~$x$
in the basis~$\acco{1,x,…,x^{\deg f-1}}$ of the quotient ring~$k[x]/f(x)$.

\medbreak

We also recall Hensel's lemma~\cite[II~(4.6)]{neukirch1999algebraic},
which is a powerful tool for solving algebraic equations in a local ring.
Let~$R$ be a complete local ring with maximal ideal~$\fr m$ and quotient
field~$k = R/\fr m$; let~$f ∈ R[x]$ be a polynomial and~$a ∈ k$ such
that~$f(a) = 0$ and~$f'(a) ≠ 0$ (\emph{i.e.} $a$~is a simple root of~$f$
modulo~$\fr m$). Then there exists a unique simple root~$b$ of~$f$ in~$R$
such that~$b ≡ a \pmod{m}$. Moreover, the computation of~$b$ is done with
Newton's approximation algorithm: $O(n)$~operations in the field~$R$
compute~$b$ up to precision~$2^n$.

%>>>1
\section{The singular part of bilinear pencils}%<<<1
\label{S:bil-sing}
\subsection{Regular and singular pencils}%<<<2
Let~$b = (b_{λ})$ be a bilinear pencil over the space~$V$. The
\emph{characteristic polynomial} of the symmetric pencil~$(b_{λ})$ is
either the polynomial~$f(λ) = \det (λ b_{∞} + b_{0})$, or its homogeneous
form~$f(λ: μ) = \det (λ b_{∞} + μ b_{0})$. If $\dim_{k} V = n$, then
$f(λ: μ)$~is homogeneous of degree~$n$ (and possibly zero).
The pencil~$(b_{λ})$ is called \emph{regular}
if the characteristic polynomial is not zero,
and \emph{singular} otherwise.
We solve the isomorphism problem for regular
pencils in section~\ref{S:bil-regular} below.

In this section, we reduce to the regular case by proving that
the singular part of a symmetric pencil is reducible to
the canonical form of Kronecker.
This form is described in the tome of Gantmacher~\cite[XII(56)]{Gantmacher};
however, the proof given there only applies to pencils over~$ℂ$,
as it uses the computation of square roots of matrices
via analytic interpolation on the spectrum.
We give here an algorithmic proof that applies to any field~$k$.
This proof is also true in characteristic two for alternating pencils,
which is the case for the polar of a quadratic pencil.

\subsection{The Kronecker decomposition} %<<<2
\label{SS:Kronecker-reduction}
% Définition: indice minimal%<<<
The pencil~$(b_{λ})$ defines a symmetric $k[λ]$-bilinear form on the
module~$V_{λ} = V ⊗_{k} k[λ]$; if $(b_{λ})$~is singular, then this form
has a non-trivial kernel~$W$. Elements of~$W$ are called \emph{isotropic}
for~$(b_{λ})$. An element~$e = e_0 + λ e_1 + … + λ^h e_h$ is isotropic iff
\begin{equation}\label{eq:isotropic}
b_0 e_0 = 0, \qquad
b_0 e_1 + b_{∞} e_0 = 0, \quad … \quad
b_0 e_h + b_{∞} e_{h-1} = 0, \qquad
b_{∞} e_{h} = 0.
\end{equation}
A \emph{minimal isotropic vector} for~$(b_{λ})$ is one with minimal
degree~$h$; this degree is the \emph{minimal index} of~$(b_{λ})$. If
$(b_{λ})$~is regular, then the minimal index is~$+∞$.
By choosing a basis of~$W$ adapted to
the filtration of~$V_{λ}$ by the degree of polynomials,
we see that $W$~has a basis~$(w_1, …,w_r)$ such that,
if $h_i$~is the degree of the isotropic vector~$w_i$,
then $w_i, …,w_r$~generate no isotropic vector of degree~$< h_i$.
The degrees~$h_i$, with their multiplicity,
are called the \emph{minimal indices} of the pencil~$(b_{λ})$.
%>>>
\begin{prop}\label{prop:minimal-indep}%<<<
Let~$e = ∑ λ^i e_i$ be a minimal isotropic vector for~$(b_{λ})$. Then
\begin{enumerate}
\item The~$h+1$ vectors~$e_0, …, e_h$ are $k$-linearly independent.
\item The~$h$ linear forms~$b_{0} e_1, …, b_{0} e_h$ are $k$-linearly
independent.
\item For all~$i, j$, $b_{0}(e_i, e_j) = b_{∞}(e_i, e_j) = 0$.
\end{enumerate}
\end{prop}

\begin{proof}
We first prove~(ii). Assume that there exists a non-trivial linear
relation~$α_1 b_0 e_1 + … + α_h b_0 e_{h} = 0$ and define
vectors~$e'_0,…,e'_{h-1}$ by $e'_i = α_{h-i} e_0 + … + α_{h} e_i$. These
vectors satisfy the relations
\begin{equation}\label{eq:relation-e'}
\begin{split}
b_0 e'_0 &= α_h b_0 e_0 = 0, \\
b_0 e'_{i+1} + b_{∞} e'_i
  &= b_0 (α_{h-i} e_1 + … + α_h e_{i+1})
  + b_{∞} (α_{h-i} e_0 + … + α_h e_i) = 0, \\
b_{∞} e'_{h-1} &= b_{∞} (α_1 e_{0} + … + α_{h} e_{h-1}) \\
 &= -b_{0} (α_1 e_{1} + … + α_{h} e_{h}) \\
 &= 0.
\end{split}
\end{equation}
This means that~$(e'_0+…+λ^{h-1} e'_{h-1})$ is isotropic and of degree~$≤
h-1$ for~$b_{λ}$, which contradicts the minimality of~$e$.

To prove~(i), let~$α_0 e_0 + … + α_h e_h = 0$ be a non-trivial linear
relation. Then since $α_1 b_0(e_1) + … + α_h b_0(e_h) = 0$, by~(ii) we
must have~$α_1 = … = α_h = 0$, which in turn implies~$e_0 = 0$. However,
in this case we see that~$e_1+…+ λ^{h-1} e_h$ is isotropic of degree~$≤
h-1$.

We now prove~(iii). For all~$i, j$, note that we have
\begin{equation}\label{eq:zero-bij}
b_{∞}(e_i,e_j) = -b_0(e_i, e_{j+1}) = -b_0(e_{j+1}, e_i) =
  b_{∞} (e_{j+1}, e_{i-1}) = b_{∞} (e_{i-1}, e_{j+1}).
\end{equation}
From this and the fact that~$b_{∞} (e_i, e_h) = b_{∞} (e_h, e_i) = 0$
and~$b_0 (e_i, e_0) = b_0 (e_0, e_i) = 0$, we deduce that for all~$0 ≤
i,j ≤ h$, we have~$b_{∞} (e_i, e_j) = b_0 (e_i, e_j) = 0$.
\end{proof}
%>>>

A \emph{Kronecker module} is a vector space~$V$ with a symmetric
pencil~$(b_{λ})$ such that the coordinates~$e_i$ of a minimal isotropic
vector~$∑ λ^i e_i$ span a space~$E$ satisfying~$E = E^{⟂}$.

\begin{prop}\label{prop:kronecker-split}%<<<
Let~$(b_{λ})$ be a symmetric pencil with minimal isotropic vector~$e =
e_0 + … + λ^h e_h$. Then $V$~has, as an orthogonal direct factor,
a Kronecker module~$K_E$ containing the vectors~$e_i$.
\end{prop}

\begin{proof}
Let $E ⊂ V$~be the subspace spanned by the vectors~$e_i$ and~$E^{⟂}$ be
its orthogonal. Prop.~\ref{prop:minimal-indep} shows that
$\dim E = h+1$, 
$\dim V/E^{⟂} = h$, %ce point n'est pas immédiat pour moi. Gilles
 and~$E ⊂ E^{⟂}$. We show
that~$E^{⟂}/E$~is an orthogonal direct factor of~$V$; its orthogonal
supplement will be the required Kronecker module.
\begin{equation}\label{eq:diagram}
\xymatrix@R-=3ex{
E\ar[r]\ar@{=}[d] \ar@<1ex>@/^/@{<.}[r]^{u} & E^{⟂}\ar[r]\ar[d]
  & E^{⟂}/E \ar[d] \\
E\ar[r] & V\ar[r]\ar[d] & V/E\ar[d]\\
 & V/E^{⟂} \ar@{=}[r] \ar@<1ex>@/^/@{.>}[u]^{v} & V/E^{⟂} \\
}\end{equation}
A split extension~$0 → K_E → V → E^{⟂}/E → 0$ is given by a
retraction~$u$ of the injection~$E ↪ E^{⟂}$ and a section~$v$ of the
projection~$V → V/E^{⟂}$; this extension is orthogonal if, for
all~$x ∈ E^{⟂}, y ∈ V/E^{⟂}$, $(x-u(x)) ⟂ v(y)$.

Write~$u(x) = ∑ u_i(x) e_i$ where~$u_0,…,u_h ∈ (E^{⟂})^{∨}$ and let the
section~$v$ be defined by elements~$v_1,…,v_h ∈ V$ such that~$b_0(e_i,
v_j) = 1$ if~$i = j$ and~$0$ otherwise.
The orthogonality condition then becomes
\begin{equation}\label{eq:orth2}
b_0 (v_j,x) = u_j(x), \quad b_{∞} (v_j,x) = -u_{j-1}(x), \qquad
\text{for all~$x ∈ E^{⟂}$, $j = 1,…,h$.}
% \text{(as elements of~$(E^{⟂})^{∨}$) for~$i=1,…,d$.}
\end{equation}
These relations uniquely determine~$u_0$ and~$u_h$, and
solutions~$(u_1,…,u_{h-1})$ exist iff the values~$v_i$ also satisfy the
relations~$b_{0} (v_{j},x) = -b_{∞} (v_{j+1},x)$ for~$j = 1,…,h-1$ and~$x
∈ E^{⟂}$.

Define a map~$∂_h: (E^{⟂})^h → ((E^{⟂})^{∨})^{h-1}$ by
\begin{equation} \label{eq:deriv-surj}
∂_h (v_1,…,v_h) = (b_{0}(v_1) + b_{∞}(v_2),…,b_{0} (v_h) + b_{∞}
(v_{h-1})).
\end{equation}
The elements of the cokernel of~$∂_h$ are exactly the isotropic vectors of
degree~$≤ h-1$ in~$E^{⟂}$; since $b$~has minimal index~$≥ h$, the
map~$∂_h$ is surjective. This proves that the map $V^h → (V/E^{⟂})^h ⊕
((E^{⟂})^{∨})^{h-1}$ defined by the relations between the~$v_j$ is
surjective, and therefore that suitable~$v_j$ exist. This proves the
orthogonality of the  decomposition~$V = K_E ⊕ (E^{⟂}/E)$.
\end{proof}%>>>

Define matrices~$K'_{h}$ of size~$(h+1) × h$
and~$K_{h}$ of size~$(2h+1) × (2h+1)$ by
\begin{equation}\label{eq:def-K}
{\def\arraystretch{.9}
K'_h = \mat{λ&&0\\1&\sddots{λ}&\\&\sddots{λ}&λ\\0&&1}}, \quad
K_{h} = \mat{0&K'_{h}\\\transpose{K'_{h}}&0}.
\end{equation}

\begin{prop}\label{prop:kronecker-matrix}%<<<
Let~$(V, b_{λ})$ be a Kronecker module with minimal index~$h$. Assume
that either~$2 ≠ 0$ in~$k$ or that $b$~is alternating. There
exists a basis of~$V$ in which the pencil~$(b_{λ})$ has the
matrix~$K_{h}$.
\end{prop}

Note in particular that the case~$h = 0$ corresponds to the matrix~$K_0$,
which is the zero matrix of size~$1 × 1$, and to a vector belonging to
all the kernels of~$b_{λ}$.

\begin{proof}
Let~$e_0 + … + λ^{h} e_h$ be a minimal isotropic vector for~$(b_{λ})$ and
$E$~be the span of the~$e_i$; we need to prove that $E$~has a supplement
which is self-orthogonal for the pencil~$(b_{λ})$. Such a supplement
corresponds to a retraction~$w$ of~$V ↪ E$ such that, for all~$x,y ∈ V$,
$(x - w(x)) ⟂ (y-w(y))$; given that~$E ⟂ E$, this amounts to
\begin{equation}\label{eq:kro-self-orth}
b_{λ} (x, y) = b_{λ} (x, w(y)) + b_{λ} (w(x), y)
\quad \text{for all~$λ$ and for $x,y ∈ V$.}
\end{equation}
A basis of~$V/E^{⟂}$ is given by vectors~$f_1,…,f_d ∈ V$ such that
$b_0(e_i, f_j) = 1$ if~$i = j$ and $0$~otherwise. Since $E = E^{⟂}$, the
family~$e_0,…,e_h;f_1,…,f_h$ is a basis of~$V$. Write~$w(f_j) = ∑ w_{ij}
e_i$. The equations~\eqref{eq:kro-self-orth} then amount to
\begin{equation}\label{eq:wij}
w_{i,j} + w_{j,i} = b_{0} (f_i, f_j);\quad
w_{i-1,j} + w_{j-1,i} = -b_{∞} (f_i, f_j).
\end{equation}
If~$2 ≠ 0$ in~$k$, then this defines the values~$w_{i,i} = \frac{1}{2} b_{0} (f_i, f_i)$
and~$w_{i-1,i} = -\frac{1}{2} b_{∞} (f_i, f_i)$; else if $b$~is
alternating, then $w_{i,i}$ and~$w_{i-1, i}$ may take any value in~$k$.
All the other coefficients~$w_{i,j}$ then follow from the
relation~$w_{i,j} - w_{i-1,j+1} = b_{0} (f_i, f_j) + b_{∞} (f_i,
f_{j+1})$.
\end{proof}
%>>>

From Props.~\ref{prop:kronecker-split}, \ref{prop:kronecker-matrix} and
an induction step on the minimal index of the pencil we deduce the
following.

\begin{prop}[Kronecker decomposition]\label{prop:kronecker}%<<<
Let~$(b_{λ})$ be a symmetric pencil on~$V$; if $2 = 0$ in~$k$, further
assume that $(b_{λ})$~is alternating.

There exists a finite sequence of integers~$(n_h)$ and 
an orthogonal isomorphism~$V ≃ ⨁ K_h^{n_h} ⊕ V'$,
where $K_h$~is the Kronecker module of index~$h$
and the restriction of~$(b_{λ})$ to~$V'$ is regular.
% Let~$(b_{λ})$ be a symmetric (or, if $2 = 0$, alternating) pencil on~$V$.
% There exists a basis of~$V$ in which the pencil has a block-diagonal
% matrix with diagonal blocks~$(K_{h_1}, …, K_{h_r}, B')$, where $K_h$~is
% the square matrix of size~$2h+1$ defined by equation~\eqref{eq:def-K},
% the integers~$h_1 ≤ … ≤ h_r$ are the minimal indices of~$(b_{λ})$, and
% $B'$~is the matrix of a regular pencil.
\end{prop}

We note that this result extends to general fields
the classical result over the real numbers~\cite[XII, §4]{Gantmacher}.
%>>>
%>>>1
\section{Linear equivalence of regular bilinear pencils}%<<<1
\label{S:bil-regular}

We give here an algorithm for computing an isomorphism between two
regular bilinear pencils. Assume that $\bm{b} = (b_{λ})$~is regular,
which means that its characteristic polynomial~$f(λ) = \det (b_0 + λ
b_{∞})$ is not zero. Then, for any~$λ$ such that~$f(λ) ≠ 0$, the bilinear
form~$b_{λ}$ is regular.

\subsection{Localisation of regular pencils}%<<<2
\label{SS:bil-reg-local}

We first prove that we may assume that one of the bilinear
forms~$(b_{λ})$ is regular.
Note that when $k = \F_q$~is a finite field, it may happen that $λ^q μ -
λ μ^q$ divides~$f(λ:μ) ≠ 0$, so that~$f(λ) = 0$ for all~$λ ∈ ℙ^1(k)$. In
this case, although $(b_{λ})$~is a regular pencil, all forms~$b_{λ}$ are
degenerate. However, the decomposition given by
Lemma~\ref{lem:decomp-bezout} below still applies.

We first isolate the subspace where $b_{∞}$~is not regular.
If $b_{0}$~is regular, then this is the subspace where
the endomorphism~$b_{0}^{-1} b_{∞}$~is nilpotent.
To make the proof work in the general case,
we replace this endomorphism by the relation~$≻$ below.

\begin{lem}\label{lem:finite}%<<<
Let~$\bm{b}$ be a symmetric pencil on~$V$. For any two vectors
$x,y ∈ V$, we write~$x ≻ y$ if~$b_{∞} x + b_0 y = 0$.
\begin{enumerate}
\item Let~$W$ be the set of~$x ∈ V$ such that there exists a chain~$x ≻ …
≻ 0$. For all~$y ∈ V$, if $b_0(y, W) = 0$ then $b_{∞}(y, W) = 0$.
\item Let~$W'$~be the $b_0$-orthogonal of~$W$. Then $b_0(W') ⊂ b_{∞}
(W')$.
\end{enumerate}
Further assume that $\bm{b}$~is regular; this means that there exists no
non-trivial chain~$0 ≻ x_0 ≻ … ≻ x_h ≻ 0$.
\begin{enumerate}
\setcounter{enumi}{2}
\item The space~$V$ decomposes as the orthogonal direct sum~$V = W ⊕ W'$.
\item The restriction of~$b_{∞}$ to~$W'$ and the restriction of~$b_{0}$
to~$W$ are injective.
\end{enumerate}
% There exists an
% orthogonal direct sum decomposition~$V = W ⊕ W'$ such that $b_{0}$~is
% injective on~$W$, and $b_{∞}$~is injective on~$W'$.
\end{lem}

\begin{proof}
(i) Assume that $b_0(y, W) = 0$ and let~$x_0 ∈ W$,
so that there exists a chain~$x_0 ≻ x_1 ≻ … ≻ 0$.
Then since $x_1 ∈ W$, we have~$b_{∞} (y, x_0) = -b_0 (y, x_1) = 0$.

(ii) Let~$\chk{V}$ be the dual space of~$V$. The space~$b_{λ}(W')$ is the set
of linear forms which are zero on all elements~$x$ such that~$b_{λ} (x,
W') = 0$.
We have to prove the following: for any vector~$x$, $b_{∞}(x, W') =
0$ implies~$b_0 (x, W') = 0$. The relation~$b_{∞} (x, W') = 0$ means
that, for all~$y ∈ V$, $b_0(W, y) = 0$~implies~$b_{∞} (x, y) = 0$; in
other words, we have~$b_{∞} (x) ∈ b_0(W)$. This means that there
exists~$w ∈ W$ such that~$x ≻ w$, which in turn implies that~$x ∈ W$. It
follows by definition of~$W'$ that~$b_0(x, W') = 0$.

(iii) Let~$y_0 ∈ W ∩ W'$. By~(ii), there exists~$y_1 ∈ W'$ such that~$y_1
≻ y_0$; this implies that~$y_1 ∈ W$. It follows that there exists an
infinite sequence~$y = (y_i) ∈ W ∩ W'$ such that~$y_{i+1} ≻ y_i$. Since
$V$~is finite-dimensional, this family is not free. Assume that~$α_0 y_0
+ … + α_m y_m = 0$ with~$α_m ≠ 0$ and define vectors~$y'_i = α_m y_i + …
+ α_0 y_{m-i}$; then we again have~$y'_i ∈ W ∩ W'$, $y'_{i+1} ≻ y'_i$,
and~$y'_m = 0$. This implies that~$0 ≻ y'_{m-1} ≻ … ≻ y'_0 ≻ 0$. Since
$\bm{b}$~is regular and~$α_m ≠ 0$, we deduce that~$y_0 = 0$ and that~$W ∩
W' = 0$. Since $\dim W' = n - \dim b_0(W) ≥ n - \dim W$, this proves
that~$W ⊕ W' = V$.

(iv) Let~$x ∈ W$ such that~$b_0(x) = 0$. Then~$0 ≻ x$. Since $\bm{b}$~is
regular, this implies~$x = 0$. 
Let~$y ∈ W'$ such that~$b_{∞} (y) = 0$. This means that~$y ≻ 0$, and
therefore~$y ∈ W$. By~(iii), this implies~$y = 0$.
\end{proof}%>>>
\begin{lem}\label{lem:decomp-bezout}%<<<
Let~$\bm{b}$ be a regular symmetric pencil on the vector space~$V$.
Let~$f(λ: μ) = \det (λ b_{∞} + μ b_0)$~be the homogeneous characteristic
polynomial of~$\bm{b}$, and let~$f = ∏ g_i$ be a factorisation of~$f$ in
mutually coprime factors.

Then there exists a unique decomposition~$V = ⨁ V_i$ such that the
spaces~$V_i$ are pairwise orthogonal for all forms of~$\bm{b}$ and the
restriction~$\bm{b}|_{V_i}$ has characteristic polynomial~$g_i$.
\end{lem}

\begin{proof}
By Lemma~\ref{lem:finite}, there exists
a unique orthogonal decomposition~$V = V_{∞} ⊕ V'$
such that $b_{∞} | V'$ and~$b_{0} | V_{∞}$~are regular
and $b_{0}^{-1} b_{∞} | V_{∞}$~is nilpotent.
The space~$V_{∞}$ corresponds to the largest power of~$μ$ dividing~$f(λ:μ)$.

Replacing~$V$ by~$V'$, we may assume that~$b_{∞}$~is regular.
This implies that $b_0$~has an adjoint endomorphism~$c = -b_{∞}^{-1} b_0$
such that~$b_0(x,y) = -b_{∞}(x, c y) = -b_{∞}(c x, y)$;
in particular, all elements of the algebra~$k[c]$
are self-adjoint with respect to~$b_{∞}$.

Let~$f(λ) = f(λ: 1)$ be the affine characteristic polynomial.
It is enough to prove the result for the decomposition~$f = gh$
where $g, h$ are mutually prime.
Let~$u, v$ be polynomials such that~$ug + vh = 1$,
and $x, y ∈ V$ such that~$g(c)(x) = 0$ and~$h(c)(y) = 0$;
we may then write
\begin{equation}\label{eq:bezout}
\begin{split}
b_{∞} (x, y) & = b_{∞} (x,\: u(c) g(c) y + v(c) h(c) y ) \\
&= b_{∞} (u(c) g(c) x, y) \,+\, b_{∞} (x, v(c) h(c) y) \\
&= 0.
\end{split}
\end{equation}
Since~$y' = c(y)$ also verifies~$h(c)(y') = 0$, equation~\eqref{eq:bezout}
also proves that~$b_{0}(x,y) = b_{∞}(x,y') = 0$, and hence~$x, y$ are
orthogonal for all forms~$b_{λ}$.
\end{proof}%>>>

The decomposition of~$V$ obtained by applying
Lemma~\ref{lem:decomp-bezout} to the full factorisation of~$f$
over~$k[x]$ is the \emph{primary decomposition} of the pencil~$(b_{λ})$.
The restriction of the pencil to each summand~$V_i$ has as its
characteristic polynomial a power of an irreducible polynomial;
such a pencil is called \emph{local}.

If two regular pencils~$\bm{b}, \bm{b}'$ are isomorphic (in the IP1S sense),
then they have the same characteristic polynomial,
and computing an isomorphism between~$\bm{b}$ and~$\bm{b}'$
is the same as computing it on each factor of the primary decomposition.
Therefore, in what follows, we shall assume that
both pencils are local (and hence finite).

\subsection{Symmetric forms commuting with a local algebra}%<<<2
\label{ss:commute}

Let~$b_{λ} = λ b_{∞} + b_0$ be a local pencil on~$V$, with characteristic
endomorphism~$c = - b_{∞}^{-1} b_0$.
Defining $R = k[c]$ makes~$V$ into a $R$-module, and we see that the
$k$-bilinear form~$b_{∞}$ on~$V$ \emph{commutes with~$R$} in the
following sense: for all~$a ∈ R$, we have~$b_{∞} (a x, y) = b_{∞} (x, a
y)$.
Moreover, the morphisms of pencils preserving the characteristic
polynomial are exactly the $R$-linear maps preserving this $k$-bilinear
form.

Since $b_{λ}$~is local, we know that
the minimal polynomial of~$c$ is of the form~$f^ℓ$ where $f$~is irreducible.
In particular, since $k$~is a finite field,
this implies that $f$~is a separable polynomial.
Let~$K$ be the extension field~$k[x] / f(x)$.
We note that $c$~is an approximate root of~$f$ in the (complete) local
algebra~$R$, so that, by Hensel's lemma, $R$~contains an exact root~$x'$
of~$f$; this turns $R$~into a $K$-algebra isomorphic to~$R_ℓ = K[π]/π^ℓ$.

\begin{prop}\label{prop:trace-form}%<<<
Let~$K/k$ be a separable field extension, and~$V$ be a finite-dimensional
vector space over~$K$.

For any $k$-bilinear form~$b: V ⊗ V → k$ commuting with~$K$, there
exists a unique $K$-bilinear form~$b_K: V ⊗ V → K$ such that~$b =
\Tr_{K/k} \,∘ \,b_K$.
\end{prop}

\begin{proof}
We first recall the (classical) proof of the result when~$V = K$.
In this case, let~$z$ be a primitive element of~$K$ and write $d =
[K:k]$. Since~$K/k$ is separable, the trace form is
non-degenerate~\cite[VI~5.2]{lang-algebra} and there exists a unique
element~$a ∈ K$ such that $\Tr (a z^i) = b(1, z^i)$ for all~$i=0, …,
d-1$.
We immadiately see that, for all~$x, y ∈ k$, $\Tr (a x y) = b(x, y)$.

The general case now follows directly from choosing a $K$-basis of~$V$:
all coordinates of a bilinear form are themselves bilinear forms.
\end{proof}%>>>

The above proposition reduces the problem to the case where~$K = k$.
For any integer~$ℓ$, we define the $k$-linear form~$τ_ℓ$ on~$R_ℓ =
k[π]/π^ℓ$ as the coefficient of~$π^{ℓ-1}$; we write~$R$ and~$τ$ instead
of~$R_ℓ$ and~$τ_ℓ$ when there is no ambiguity.
We note that $τ(xy)$, as a $k$-bilinear form on~$R$, is regular and
commutes with~$R$.
Actually, multiplication by~$π^{ℓ-1}$ defines a (non-canonical)
$R$-linear isomorphism between~$R$ and its dual (as a $k$-vector
space)~$R^{∨}$.

\begin{prop}\label{prop:trace-local}%<<<
Let~$M, N$ be $R$-modules of finite length.
For any $k$-bilinear form~$b: M ⊗ N → k$ commuting with~$R$,
there exists a unique $R$-bilinear form~$b_R: M ⊗ N → R$ such that~$b = τ
∘ b_R$.
\end{prop}

Note that we do not demand that $M$, $N$ be \emph{free} as $R$-modules.
Moreover, the unicity of~$b_R$ shows that both forms~$b$ and~$b_R$ are
simultaneously symmetric or antisymmetric.
However, when $k$~is a field of characteristic two, it may happen (as we
shall see in Section~\ref{S:quad-reg}) that $b$~is alternating whereas
$b_R$~is not.

\begin{proof}[\protect{Proof of Prop.~\ref{prop:trace-local}}]
By the structure theorem of modules over a principal ring, we may
write~$M = ⊕ R_{m_i}$ for some finite sequence of integers~$m_i$. If the
result holds for modules~$(M', N)$ and~$(M'', N)$ then it also holds
for~$(M' ⊕ M'', N)$. Therefore, using induction on the length of both
modules, we only need to prove the case where~$M = R_m$ and~$N = R_n$
where~$m ≥ n$. In this case, since $b$~commutes with~$k[π]$, we see that
for~$x = ∑ x_i π^i$, $y = ∑ y_i π^i$:
\begin{equation}
b(x,y) \;=\; ∑_{i,j} x_i y_j\: b(1, π^{i+j})
  \;=\; ∑_{i+j+r = m-1} x_i y_j\: b(1, π^{m-1-r}).
\end{equation}
Let~$a = ∑ b(1, π^{i}) π^{m-1-i}$. Since $b(1, π^{r}) = 0$ for all~$r ≥
n$, $a$~belongs to the ideal~$π^{m-n} R_m = \Hom_R (R_n, R_m)$, so that
the product~$a\,y$ is well-defined in~$R_m$. The bilinear form~$b_R$ is
finally the form defined by~$b_R(x,y) = a\:x y$.
\end{proof}%>>>

By Propositions~\ref{prop:trace-form} and~\ref{prop:trace-local},
we see that for any local pencil~$(b_{∞}, b_0)$
with characteristic endomorphism~$c$
and associated local ring~$R = k[c]$,
there exists a unique $R$-bilinear form~$b_R$ on~$V$
such that~$b_{∞} = \Tr_{K/k} ∘ τ_ℓ ∘ b_R$.
Local pencils and linear morphisms are thus equivalent
to $R$-bilinear forms on~$V$ and $R$-linear maps.
% Moreover, a morphism of pencils~$s: b → b'$ is a
% $k$-linear map such that~$\transpose{s} ∘ b_{∞} ∘ s = b'_{∞}$
% and~$s^{-1} ∘ c ∘ s = c'$, or in other words a $R$-linear congruence
% between the $R$-bilinear forms~$b_R$ and~$b'_R$. Therefore, solving the
% local IP1S problem amounts to computing a congruence matrix
% between two~$R$-bilinear forms on a $R$-module of finite length.

\subsection{Classification of bilinear forms over a local algebra}%<<<2

We conclude the proof of the odd-characteristic case with the
classification of bilinear forms over a local algebra.
We provide slightly adapted proofs of the classical
theory~\cite{milnorhusemoller,omeara}.
The result over the local algebra is a variant of
the classification over the (finite) residue field;
the reduction algorithm itself is a version of Gauß' reduction algorithm
for quadratic forms.

\begin{prop}\label{prop:diag-bigblock}%<<<
Let~$M$ be a $R$-module of finite length and $b$~be a $R$-bilinear form
on~$M$.
If the $k$-bilinear form~$τ ∘ b$ is regular,
then there exists an orthogonal decomposition~$M = ⨁ M_m$,
where for $m ≤ ℓ$, $M_m$~is a finite free module over~$R_m = R / π^m$
and $b$~is regular, as a $R_m$-bilinear form, on each module~$M_m$.
\end{prop}

\begin{proof}
We reason by induction on the length~$ℓ$ of~$R$. By the structure theorem for
modules over the principal ring~$R$, there exists a decomposition~$M = F
⊕ N$ where $F$~is free over~$R = R_ℓ$ and~$π^{ℓ-1} N = 0$. In
particular, $N$~is a $R'$-module, where~$R' = R_{ℓ-1}$; we define~$τ'
= τ_{ℓ-1}: R' → k$ and note that~$τ'(a) = τ(π a)$ for all~$a ∈ R'$.

We first show that the restriction of~$b$ to~$F$ is regular. Let~$x ∈ F$
such that~$R x$ is a direct factor: this means that~$π^{ℓ-1} x ≠ 0$.
Since $τ ∘ b$~is regular, there exists~$y ∈ M$ such that~$τ(b(π^{ℓ-1} x,
y)) = 1$. This implies that~$b(x, y) ≡ 1 \pmod{π}$. Let~$y = y' + y''$
where~$y' ∈ F$ and~$y'' ∈ M/F$: since~$π^{ℓ-1} y'' = 0$, we have~$π^{ℓ-1}
b(x, y'') = 0$ and therefore~$b(x, y'') ∈ π R$. Therefore, $a = b(x, y')
≡ b(x, y) ≡ 1 \pmod{π}$. In particular, $a ∈ R^{×}$, so that~$b(x, a^{-1}
y') = 1$, which shows that $b$~is regular on~$F$.

Let now~$y ∈ N$. Since $b$~is regular on~$F$, there exists a unique~$f(y)
∈ F$ such that, for all~$x ∈ F$, $b(x, y) = b(x, f(y))$. This implies
that the map~$N → M, y ↦ y - f(y)$ is orthogonal to~$F$, and therefore
defines a $b$-orthogonal decomposition~$M = F ⊕ N$. Finally,
since~$π^{m-1} y = 0$, the map~$b$ has its values in~$π R$, which means
that there exists a $R'$-bilinear form~$b'$ on~$N$ such that~$b = π b'$;
since $τ ∘ b$~is regular, its orthogonal summand $τ' ∘ b'$~is regular,
and we may apply the induction hypothesis to the $R'$-bilinear form~$b'$
on~$N$.
\end{proof}%>>>

We shall generally call $b$ \emph{regular} if $τ ∘ b$~is regular as a
$k$-bilinear form. If $M$~is free then this does not conflict with the
standard definition of $R$-regularity.

Note that, if we do not assume $b$~to be regular, then there does not
necessarily exist a decomposition equivalent to that of
Prop.~\ref{prop:diag-bigblock}. For example, the $R_2$-bilinear form
over~$M = R_1 ⊕ R_2$ defined by~$b(x_1 ⊕ x_2, y_1 ⊕ y_2) = (π x_1) y_2 -
x_2 (π y_1)$ is not diagonalizable.

\begin{prop}\label{prop:bilinear-odd}%<<<
Assume that $k$~is finite and~$2 ≠ 0$; let~$Δ$ be a non-square element
of~$k^{×}$.
Then any regular symmetric $R$-bilinear form~$b$ on a free $R$-module~$M$
is equivalent to one of the diagonal forms~$(1, …, 1)$ or~$(1, …, Δ)$.
\end{prop}

\begin{proof}
The proof follows from applying a Hensel lift to the classical proof over
finite fields~\cite[IV(1.5)]{milnorhusemoller}.
By the Gram orthogonalization algorithm~\cite[I(3.4)]{milnorhusemoller},
$b$~is congruent to a bilinear form with diagonal matrix~$A =
\mathrm{diag} (a_1, …, a_n)$. Since $x^2$~is a separable
polynomial over~$R$, two applications of Hensel's lemma in the complete
ring~$R$ allow the following lifts to~$R$ of results in the finite
field~$k$:
\begin{enumerate}
\item[(a)] for all~$i$, we have either~$a_i = b_i^2$ or~$a_i = b_i^2 Δ$
in~$R$;
\item[(b)] the equation~$u^2 + v^2 = Δ$ has a solution with~$u, v ∈ R$.
\end{enumerate}
By~(a), we may assume that~$a_i = 1$ or~$a_i = Δ$. From~(b), we deduce
the matrix relation
\begin{equation}
\transpose{\mat{u&-v\\v&u}}·\mat{1&0\\0&1}·\mat{u&-v\\v&u} =
\mat{Δ&0\\0&Δ}.
\end{equation}
This allows canceling all pairs of~$Δ$ appearing in the diagonalization
of~$A$.
\end{proof}%>>>

\subsection{Solving the general case of IP1S}%<<<2
\begin{thm}\label{thm:IP1S}
Let~$k$ be a finite field of characteristic~$≠2$ and~$(b_{λ})$ be a
pencil of $n$-dimensional symmetric bilinear forms over~$k$.
It is possible, using no more than~$\Ot(n^3(h+1)) ≤ \Ot(n^4)$ operations
in~$k$, where $h ≤ n$~is the largest of the minimal indices of~$(b_{λ})$,
to compute an isomorphism between~$(b_{λ})$ and a (unique) block-diagonal
pencil with diagonal blocks of the following form:
\begin{enumerate}
\item \emph{Kronecker blocks}~$K_{h} =
\mat{0&K'_{h}\\\transpose{K'_h}&0}$ for integers~$h ≥ 0$, as defined in
Prop.~\ref{prop:kronecker};
\item \emph{finite local blocks}~$L_{f,ℓ,u}$, defined as the~$ℓ ×
ℓ$-block matrix
\[ L_{f,ℓ,u} \;=\; \mat{0&&-T_f u&T_f u (λ - M_f)\\
  &\siddots{-T_f u}&\siddots{-T_f u}&\\
  -T_f u&\siddots{-T_f u}&\\T_f u(λ-M_f)&&&0}, \]
where $f$~is an irreducible polynomial, $M_f$~is the companion matrix
of~$f$, $T_f$~is a prescribed invertible matrix such that both~$T_f$
and~$T_f M_f$~are symmetric, $ℓ$~is an integer, and $u$~is either the
identity matrix or a prescribed non-square element of the field~$k[M_f]$,
with the extra condition that for fixed~$(f, ℓ)$, at most one of the
values~$u$ may be different from~$1$;
\item \emph{infinite local blocks}~$L_{∞,ℓ,u}$, defined as
the~$ℓ×ℓ$-matrix
\[ L_{∞,ℓ,u} \;=\; u\mat{0&&-λ&1\\&\siddots{-λ}&\siddots{-λ}&\\
  -λ&\siddots{-λ}&&\\1&&&0}, \]
where $ℓ$~is an integer, $u$~is either 1 or a prescribed non-square
element of~$k$, and for fixed~$ℓ$, at most one of the values~$u$ may be
different from~$1$.
\end{enumerate}
\end{thm}

This theorem solves the IP1S problem in time~$O(n^4)$: given two
pencils~$\bm{a}$ and~$\bm{b}$, we transform both of them to
the canonical form above. This form will be the same iff the two pencils
are isomorphic in the IP1S sense, and in this case, composing the two
transformations gives an answer to the computational IP1S problem.

\paragraph{Algorithm and complexity.}

The algorithm corresponding to Theorem~\ref{thm:IP1S} for a
pencil~$\bm{b}$ decomposes in the three following steps.

\begin{enumerate}
\item \label{it:alg-kronecker} Compute the Kronecker decomposition: as
long as the kernel of the matrix~$b_{λ}$ is not trivial, compute a
minimal isotropic vector~$e = ∑ λ^i e_i$ and the according Kronecker
block as an orthogonal direct factor, according to
Prop.~\ref{prop:kronecker-split}.
\item \label{it:alg-factor} Now~$\bm{b}$~is regular. Compute and factor
its characteristic polynomial~$f(λ)$ and split~$V$ as an orthogonal direct
sum of primary components~$V_f$ for each prime divisor~$f$.
On the the “infinite” factor corresponding to the divisor~$μ$ of~$f(λ:μ)$,
swap the forms~$b_{∞}$ and~$b_0$.
\item \label{it:alg-local} For each prime divisor~$f$, write the local
pencil~$\bm{b}|V_f$ at~$f$ as a matrix with entries in~$K = k[x]/f(x)$.
Perform the reduction of~\ref{prop:diag-bigblock} and write~$\bm{b}|V_f$
as an orthogonal direct sum of quadratic forms over algebras~$K[π]/π^{ℓ}$,
and then reduce each of these forms to
one of the two canonical diagonal forms.
\end{enumerate}

Most of the linear algebra steps, including computing the
rational normal form of the regular part of the pencil, may be done
in~$\Ot(n^3)$ field operations~\cite{kaltoffen11compute}.
In particular, computing the  Frobenius rational normal form
includes factoring the characteristic polynomial.
This factorization may also be done, again in cubic time,
using a dedicated factoring algorithm.
Reduction of quadratic forms over the local algebras is just
reduction over the residue field
(which uses a square root computation in this finite field),
followed by a Hensel lift.

The only step not covered by standard algorithms is the reduction to
Kronecker normal form performed in Step~\itemref{it:alg-kronecker}.
Computing the minimal isotropic vectors requires solving
a chain of~$h$ linear equations of the form~$b_{∞}(x) = b_0(y)$
and hence has a complexity~$O(n^3(h+1))$.
The same applies to the computation of a preimage by the map~$∂_h$
of~\eqref{eq:deriv-surj}.
\ifapx This algorithm is detailed in Appendix~\ref{A:algo} below. \fi

\paragraph{Remarks.}

The only place where the finiteness of~$k$ is required in the proof of
this theorem is for the structure of quadratic forms over~$k$ in
Prop.~\ref{prop:bilinear-odd}. Even when $k$~is infinite, if it is perfect
and has a good theory of quadratic forms,
we expect it to translate to a good theory of the IP1S problem over~$k$.

We also note that, if $(b_{λ})$~is regular,
then~$h = 0$ and the algorithm in this case has complexity~$\Ot(n^3)$. As
explained before, this is the dominant case and we therefore expect any
implementation on random pencils to run in average time~$\Ot(n^3)$.

There exist cubic algorithms computing the Kronecker decomposition of pencils of
linear maps over a characteristic zero field~\cite{beelen1988improved}.
These algorithms are not directly applicable over a finite field as they
use some rotations over the real numbers and are mostly concerned with
numerical stability; more importantly, they work with linear maps up to
equivalence, whereas we need quadratic forms up to congruence. However, as the
corresponding problem over a finite field has not been much studied, the
existence of a faster algorithm for computing the Kronecker decomposition is not
unlikely.

\paragraph{Comparison to the “polar decomposition” algorithm}

Berthomieu et al. suggested a “polar decomposition” algorithm for
the IP1S problem~\cite{DBLP:journals/corr/BerthomieuFP13}.
This algorithm, inspired by real analysis techniques,
decomposes in two steps.
Let~$D$ be a regular invertible matrix and $H$~be any matrix.
We recall that the $D$-adjoint of a matrix~$M$
is~$M^{⋆} = D^{-1} · \transpose{M} · D$.
The “polar decomposition” algorithm makes the two following claims.
\begin{enumerate}
\item[(A)] Let~$Y$ be a regular matrix commuting with both~$H$ and~$H^{⋆}$.
Then the matrix~$Z = D· Y · \transpose{Y} · D^{-1}$ has a square root~$W$.
\item[(B)] Define $X = Y · W^{-1}$. Then $X$~is a solution to the IP1S
problem $H· X = X · H$ and~$X^{⋆} · X = 1$.
\end{enumerate}
We give here counter-examples to both claims above
over any finite field with odd characteristic.
In the IP1S case, $D$~is the matrix of the bilinear form~$b_{∞}$, and
$H$~is the matrix of the characteristic endomorphism~$b_{∞}^{-1} b_0$.
This gives an extra condition, not used
in~\cite{DBLP:journals/corr/BerthomieuFP13}:
namely, since $b_0$~is symmetric, we must have~$H^{⋆} = H$.
% We note that the authors of~\cite{DBLP:journals/corr/BerthomieuFP13}
% further assume that $D$~is diagonal (they show that the general case is
% reducible to this one).

\medskip

For claim A, let~$d$ be a non-square element of~$k$.
Since $k$~is finite, there exist~$u, v ∈ k$ such that~$d = u^2 + v^2$.
Let~$t ∈ k$ such that~$t^2 ≠ 1$ and define
\[ D = \mat{t & 0 \\ 0 & 1}, \quad
H = \mat{0 & 1 \\ t & -\frac{u}{v}(t+1)}. \]
We then see that~$H^{⋆} = H$ and that
the matrix~$Y = \mat{u & v \\ t v & -t u}$ commutes with~$H$.
However, we easily check that
$Z = D · Y · \transpose{Y} · D^{-1} = \mat{d & 0 \\ 0 & t^2 d}$~has no square
root.

Since the only hypothesis on~$d$ is that it is a sum of squares,
this counterexample shows that the “polar decomposition” algorithm
requires a (tower of) \emph{real quadratic extensions} of the base field,
where a real quadratic extension is obtained by adjoining
the square root of a sum of squares.
The fields having no non-trivial real quadratic extensions
are the \emph{Euclidean fields};
this includes quadratically closed fields
and real closed fields such as~$ℝ$ or~$\overline{ℚ} ∩ ℝ$,
but since any element in a finite field is a sum of (two) squares,
no finite field is Euclidean.

\medskip

For claim B, let~$d$ be any element of~$k$ and again write~$d = u^2 + v^2$.
We keep the first example but now use the case where~$t = -1$: let
\[ D = \mat{-1 & 0 \\ 0 & 1}, \quad
H = \mat{0 & 1 \\ -1 & 0}. \]
We again have $H^{⋆} = H$, and $Y = \mat{u & v\\-v & u}$~commutes
with~$H$.
However, $Z = D · Y · \transpose{Y} · D^{-1} = \mat{d & 0 \\ 0 & d} = W^2$
where $W = \mat{0 & d \\ 1 & 0}$ (or any of its conjugates).
We now easily check that
$X = Y · W^{-1}$~is not a solution of the IP1S problem.

In the real case, the “polar decomposition” method computedsthe
square root of the matrix~$Z$ by analytic interpolation on its spectrum;
this root then belongs to the algebra~$k[Z]$
and therefore commutes with any matrix commuting with~$Z$.
In the case above, since $Z = d · 1_{2}$ with~$d = u^2 + v^2 ≥ 0$,
the analytic interpolation will compute one of the roots~$±√{d} · 1_2$.
The square root~$W$ we give above does not belong to the algebra~$k[Z]$,
thus making this method fail.

\paragraph{Decisional IP1S and extensions of scalars.}

The solution of IP1S over an extension field
in~\cite{DBLP:journals/corr/BerthomieuFP13} raises the following
question: given two pencils~$\bm{a}$, $\bm{b}$ defined over a field~$k$
and IP1S-equivalent over an extension~$k'$ of~$k$, are they always
equivalent over the base field~$k$?

The structure given by theorem~\ref{thm:IP1S}
gives a simple, negative answer to this question.
Since the Kronecker blocks are invariant by extension of scalars,
the problem reduces to the finite and infinite local blocks~$L_{f,ℓ,u}$.
% For any irreducible polynomial~$f$ factoring over~$k'$ as~$f = ∏ f_i$,
% we have $L_{f,ℓ,u} ⊗_{k} k' = ∏ L_{f_i, ℓ, u'_i}$,
% where $u'_i$~is the image of~$u$
% by the projection map~$k[x]/f(x) → k'[x]/f_i(x)$.
This implies that $\bm{a}$ and~$\bm{b}$ are equivalent over~$k$
if and only if, for all irreducible factors~$f$, they have the same value
$u_{(f, ℓ)} (\bm{a}) = u_{(f, ℓ)} (\bm{b}) ∈ K_f^{×}/(K_f^{×})^2$,
where $K_f = k[x]/f(x)$~is the extension of~$k$ generated by~$f$.
Since the norm map defines
an isomorphism~$K_f^{×}/(K_f^{×})^{2} → k^{×} / (k^{×})^2$,
this condition is equivalent to: $\bm{a}$ and~$\bm{b}$ have the same
characters~$χ_{f,ℓ} (\bm{a}) = N_{K_f/k} (u_{(f,ℓ)} (\bm{a}))$.

% For an irreducible polynomial~$f$, we have
% \begin{equation}\label{local-extension}
% L_{f,ℓ,u} ⊗_{k} k' = %\begin{cases}
% %L_{p,d,u'},&\text{if $p$~is irreducible over~$k'$;}\\
% ⨁ L_{f_i, ℓ,u'}, %&
% \text{if $f$~splits as~$∏_i f_i$ over~$k'$,}
% %\end{cases}
% \end{equation}
% where $u'$~is the image in~$k'^{×}/(k'^{×})^2$ of~$u$. The
% map~$k^{×}/(k^{×})^2 → k'^{×}/(k'^{×})^2$ is bijective when $[k':k]$~is
% odd and zero when it is even. From this we deduce:
% 
% \begin{prop}\label{prop:IP1S-extension}
% Let~$\bm{a}, \bm{b}$ be two pencils which are IP1S-equivalent over a
% field extension~$k'$ of the finite field~$k$.
% \begin{enumerate}
% \item If the degree~$[k':k]$~is odd, then $\bm{a}$ and~$\bm{b}$ are
% equivalent over~$k$.
% \item If the degree~$[k':k]$~is even, them $\bm{a}$ and~$\bm{b}$ are
% equivalent over~$k$ iff, for all local factors~$L_{f,ℓ,u}$, they have the
% same value~$u(f,ℓ) ∈ k^{×}/(k^{×})^2$.
% \end{enumerate}
% \end{prop}

% In particular, when $k → k'$~has even degree,
% solving the IP1S problem over~$k'$
% does not, in general, solve the decisional form of IP1S over~$k$.
For example, let~$a ∈ k$ and define the two-dimensional
quadratic pencils~$\bm{b} = (2 x y, x^2 + a y^2)$
and~$\bm{b'} = (x^2 + y^2/a, 2 x y)$.
These two pencils are equivalent only over any extension of~$k$
containing a root of the equation~$x^4 + 4 a = 0$.
Although this fact is easy to check by hand,
we also provide an interpretation using our work.
The characteristic endomorphism of~$\bm{b}$ is
\begin{equation*}
c = -\mat{0&1\\1&0}^{-1} · \mat{1&0\\0&a} \;=\; \mat{0 & -a\\-1 & 0};
\end{equation*}
its characteristic polynomial is~$x^2 - a$.
Let~$R = k[c]$.
The space~$V = k^2$ is cyclic as a $R$-module and generated by the
vector~$e = \def\arraystretch{.5}\mat{1\\0}$,
with~$c · e = \def\arraystretch{.5}\mat{0\\-1}$.
Since $R$~is a separable $k$-algebra,
we may apply Prop.~\ref{prop:trace-form} even when it is not a field.
Let~$b_R = (x, y) ↦ b_R x y$ be the lift of~$b_{∞}$ to~$R$; then
\begin{equation}
\Tr_{R/k} b_R = b_{∞} (e, e) = 0
\quad\text{and}\quad
\Tr_{R/k} (c b_R) = b_{∞} (e, c·e) = -1,
\end{equation}
which means that~$b_R = -\frac{1}{2 a} c$.

The same computations for the pencil~$\bm{b'}$
yield~$k[c'] ≃ k[c]$ and~$b'_R = \frac{1}{2}$.
Therefore, the pencils~$\bm{b}$ and~$\bm{b'}$
are isomorphic exactly over the extensions~$k'$ of~$k$
where $b'_R/b_R = -c$~is a square.
When writing~$-c = (u/2 + v c)^{2}$,
this means that $u^4 + 4 a = 0$ as above.
Depending on the precise value of~$a$, the pencils~$\bm{b}$ and~$\bm{b'}$
may become isomorphic over an extension of degree~$1$, $2$ or~$4$ of~$k$.

%>>>1
\section{Fully singular quadratic pencils in characteristic two}%<<<1
\label{S:quad-sing}
In this section and the following one, we assume that the field~$k$ has
characteristic two.

\subsection{Quadratic and bilinear pencils}%<<<2

We write~$σ: x ↦ x^2$ for the absolute Frobenius automorphism of~$k$.

Let~$q$ be a quadratic form over a~$k$-vector space~$V$. Since $2 = 0$
in~$k$, the polarity equation shows that the associated polar form~$b$ is
an \emph{alternating} bilinear form; namely, it satisfies~$b(x,x) = 0$
for all~$x ∈ V$. Moreover, the polarity map is not a bijection from
quadratic forms to alternating bilinear forms; its kernel is the space of
\emph{$σ$-linear forms} on~$V$. This means that, if a basis of~$V$ is
chosen, then a quadratic form is determined by its polar (which is an
alternating matrix) and its diagonal coefficients (corresponding to a
$σ$-linear form).

\bigbreak

We say that a bilinear pencil is \emph{fully singular} if
its regular part, in the sense of Prop.~\ref{prop:kronecker}, is zero.
Any fully singular alternating bilinear pencil is isomorphic to
an orthogonal sum of Kronecker modules~$K_d$.
However, in characteristic two, this gives a standard form only for
bilinear pencils and not for quadratic pencils. We explain here how to
compute an isomorphism, when it exists, between two fully singular
quadratic pencils in characteristic two. For this, we give a full
description of the group of automorphisms of an orthogonal sum of
Kronecker bilinear modules, and then show how such an automorphism acts
on the diagonal coefficients of a quadratic pencil.
We conclude the proof by showing that the corresponding equations
may be solved in polynomial time.

\subsection{An intrinsic description of Kronecker modules}%<<<2
We give an intrinsic construction of the Kronecker modules.
Except where indicated, all tensor products and duals are understood as
operations on $k$-vector spaces.

For any integer~$d ≥ 0$, let~$H_d$ be
the $d+1$-dimensional vector space of homogeneous polynomials
of degree~$d$ in the variables~$(x:y)$.
We also write~$\chk{H_d}$ for the vector space dual to~$H_d$
and~$\chev{φ,f}: \chk{H_d} × H_d → k$ for the standard bilinear pairing.

For any $h ∈ H_m$, multiplication by~$h$ defines a linear map,
which we again write~$h: H_d → H_{m+d}$,
as well as a transposed map~$\chk{h}: \chk{H_{d+m}} → \chk{H_d}$.
This means that~$\chev{\chk{h} φ, f} = \chev{φ, hf}$ for all~$φ ∈
\chk{H_{d+m}}$ and~$f ∈ H_d$.
All the maps~$h$ and~$\chk{h}$ commute with each other.

The \emph{Kronecker module} of degree~$d$ is the $2d+1$-dimensional
vector space~$K_d = H_{d-1} ⊕ \chk{H_{d}}$, equipped with the symmetric
bilinear pencil~$(b_{λ})$ defined, for~$f, f' ∈ H_{d-1}$ and~$φ, φ' ∈
\chk{H_d}$, by:
\begin{equation}\label{eq:kronecker-bilinear}
b_{λ} (f, f') \;=\; b_{λ} (φ, φ') \;=\; 0; \qquad
b_{λ} (f, φ) \;=\; \chev {φ, (λ y - x) f}.
\end{equation}

The next proposition shows that multiplication by polynomials essentially
defines all homomorphisms between the bilinear spaces~$K_d$.
As in Subsection~\ref{SS:bil-reg-local},
we write~$f ≻ g$ for the relation~$b_{∞} f + b_0 g = 0$.

\begin{prop}\label{prop:hom-Kd}%<<<
For any integers~$d, d'$, the homomorphisms from~$K_d$ to~$K_{d'}$
preserving the binary relation~$≻$ are the maps of the form
\begin{equation*}
\begin{array}{lclclcl}H_{d-1} &⊕& \chk{H_d} &→&H_{d'-1} &⊕& \chk{H_{d'}}\\
(f,&& φ) & ↦ & (α f + \chk{f} γ,&& \chk{β} φ)
\end{array},
\end{equation*}
where $α ∈ H_{d'-d}$, $β ∈ H_{d-d'}$, and~$γ ∈ \chk{H_{d+d'-1}}$.
\end{prop}

\begin{proof}
We see that, for all~$f, g ∈ H_{d-1}$ and~$φ, ψ ∈ \chk{H_d}$,
\begin{equation}\label{eq:chain-Kd}
f ≻ g \text{ iff } y f = x g; \quad
φ ≻ ψ \text{ iff } \chk{y} φ = \chk{x} ψ.
\end{equation}
Let~$(x^{i} y^{d-1-i})$ be a basis of~$H_{d-1}$
and write~$ξ_{j,d-j}$ for the dual basis of~$\chk{H_d}$.
We then have the relations
\begin{equation}\label{eq:chain-Kd-zero}
0 ≻ ξ_{0,d} ≻ ξ_{1,d-1} ≻ … ≻ ξ_{d,0} ≻ 0
\end{equation}

Let~$F: K_{d} → K_{d'}$ be a $≻$-homomorphism.
For~$j = 0, …, d$, let~$g_j$ be
the projection to~$H_{d'-1}$ of~$F(ξ_{j, d-j})$.
Applying~$F$ to the relation~\eqref{eq:chain-Kd-zero}, we see that
\begin{equation}
0 ≻ g_0 ≻ g_1 ≻ … ≻ g_d ≻ 0.
\end{equation}
This and~\eqref{eq:chain-Kd} imply that~$g_j = 0$ for all~$j$.
From this we deduce that $F(\chk{H_d}) ⊂ \chk{H_{d'}}$.

Write~$β_j = \chev{F(ξ_{j,d-j}), y^{d'}}$ and~$β = ∑ β_j x^j y^{d-d'-j}$.
By~\eqref{eq:chain-Kd-zero},
we see that $β_j = \chev{F(ξ_{0,d}), x^{j} y^{d'-j}}$,
so that~$F(ξ_{0,d}) = ∑ β_j ξ_{j,d'-j} = \chk{β} ξ_{0,d}$.
Applying~\eqref{eq:chain-Kd-zero} once more,
we deduce from this that $F(ξ_{j,d-j}) = \chk{β} ξ_{j,d-j}$ for all~$j$,
and therefore~$F(φ) = \chk{β} φ$ for all~$φ ∈ \chk{H_{d'}}$.
\end{proof}%>>>
\subsection{Kronecker modules with coefficients}%<<<2

A \emph{coefficient space} is a bilinear space~$E$ isomorphic to~$k^n$,
together with its standard scalar product~$u, v ↦ u · v = ∑ u_i v_i$. For
any linear maps~$α: V → E$, $β: W → E$, we write~$α · β$ for the
corresponding bilinear form on~$V × W$.

For~$E = k^n$, the bilinear module~$K_d^{n}$ is isomorphic to~$E ⊗ K_d$.
The multiplication maps for homogeneous polynomials defines in a natural
way a bilinear action of~$E ⊗ H_m$ on~$H_d$ and~$\chk{H_d}$: namely, for
any~$u ∈ E ⊗ H_m$ written as~$u = ∑ u_i ⊗ x^i$ with~$u_i ∈ k^n$ and~$x^i
∈ H_m$ and~$f ∈ H_d$ and~$φ ∈ \chk{H_d}$, we define
\begin{equation}
u f \;=\; ∑ u_i ⊗ (x^i f) \;∈ E ⊗ H_{d+m} \quad\text{and}\quad
\chk{u} φ \;=\; ∑ u_i ⊗ (\chk{(x^i)} φ) \;∈ E ⊗ \chk{H_{d-m}}.
\end{equation}
Likewise, let~$h ∈ \Hom(E, E') ⊗ H_m$ be written as~$h = ∑ h_i ⊗ x^i$
with~$h_i ∈ \Hom (E, E')$. For any~$u ⊗ f ∈ E ⊗ H_d$, we define~$h(u ⊗
f) ∈ E' ⊗ H_{d+m}$ by~$h(u ⊗ f) = ∑ h_i(u) ⊗ (x^i f)$.
% An element
% of~$\Hom (E ⊗ H_d, E' ⊗ H_{d+m})$ defined in this way is called
% \emph{principal}. In the same way, we define principal elements of~$\Hom
% (E ⊗ \chk{H_{d+m}}, E' ⊗ \chk{H_{d}})$.
% 
% Let~$r ≤ s$ be two integers and~$E, E'$ be two coefficient spaces. We say
% that a bilinear form~$B$ on~$(E ⊗ H_r) × (E' ⊗ \chk{H_s})$ is
% \emph{principal} if it is of the form~$B(f, φ) = \chev {φ, hf} = \chev
% {\chk{h} φ, f}$ for some~$h ∈ \Hom (E, E') ⊗ H_{s-r}$.

% \begin{lem}\label{lem:principal}
% Let~$r ≤ s$ be two integers and~$E, E'$ be two coefficient spaces.
% \begin{enumerate}
% \item A bilinear form~$B$ on~$(E ⊗ H_r) × (E' ⊗ \chk{H_s})$ is principal
% if, and only if, for any polynomial~$z$, $B(zf, φ) = B(f, \chk{z} φ)$.
% \item Let~$α: E ⊗ H_r → E' ⊗ H_s$ and~$β: E' ⊗ \chk{H_s} → E ⊗ \chk{H_r}$ be
% two linear maps. If $1 · α + β · 1$~is principal, then both~$α$ and~$β$
% are principal.
% \end{enumerate}
% \end{lem}
% 
% \begin{proof}
% By taking coordinates on~$E$ and~$E'$, we reduce the proof to the case
% where~$E = E' = k$.
% \end{proof}
% \subsection{Automorphisms of a direct sum of Kronecker modules}%<<<2

For any totally irregular pencil of quadrics~$q$ on~$V$, there exists an
unique finite sequence of integers~$(n_d)$ such that $V$~is isomorphic to
the orthogonal sum~$⨁ E_d ⊗ K_d$, where $E_d = k^{n_d}$ is a coefficient
space.

\begin{prop}\label{prop:aut-Ed-Kd}%<<<
The automorphisms of~$⨁ E_d ⊗ K_d$ are exactly the maps of the form
\begin{equation}\label{eq:aut-Ed-Kd}
u ⊗ f ↦ ∑_{d' ≥ d} α_{d',d} (u) f + ∑_{d'} \chk{f} γ_{d',d} (u), \quad
u ⊗ φ ↦  ∑_{d' ≤ d} \chk{β_{d',d} (u)} φ,
\end{equation}
where~$α_{d',d} ∈ \Hom (E_d, E_{d'}) ⊗ H_{d'-d}$, $β_{d',d} ∈ \Hom (E_d,
E_{d'} ⊗ H_{d-d'})$ and~$γ_{d',d} ∈ \Hom (E_{d}, E_{d'}) ⊗
\chk{H_{d+d'-1}}$ satisfy the following relations: let~$A, B, C$ be the
matrices~$(α_{d',d})$, $(β_{d',d})$, and~$(γ_{d',d})$; then
\begin{equation}\label{eq:aut-Ed-Kd-ABC}
\transpose{B} · A = 1, \quad\text{and}\quad
\transpose{C} · A + \transpose{A} · C = 0.
\end{equation}
\end{prop}

In the above proposition, we understand the elements~$u ⊗ f$ and~$u ⊗ φ$ to
belong to the spaces~$E_d ⊗ H_{d-1}$ and~$E_d ⊗ \chk{H_d}$.
For~$d > d'$, the space~$H_{d' - d}$ is zero,
and therefore~$α_{d,d'} = β_{d',d} = 0$.

The relation~$\transpose{A} · B = 1$ means that, for all~$d, d'$,
$∑_i \transpose{α_{i,d}} · β_{i,d'} ∈ \chk{E_d} ⊗ \chk{E_{d'}} ⊗ H_{d'-d}$
is the standard scalar product of~$E_d$ if $d = d'$ and~$0$ else;
this product is to be interpreted as
the collection of the corresponding terms of all degrees in~$(x:y)$,
and the scalar product is taken in~$E_i$.
Therefore, the elements~$α_{i,j}$ uniquely determine all of the~$β_{i,j}$.
Likewise, the relation~$\transpose{A} · C + \transpose{C} · A = 0$
means that $∑_i \transpose{α_{i,d}} · γ_{i,d'} +
\transpose{γ_{i,d}} · α_{i,d'} = 0 ∈ E_d ⊗ E_{d'} ⊗ H_{d+d'-1}$,
where scalar products are taken in~$E_i$.

% \begin{proof}[(Sketch of) Proof of Prop.~\ref{prop:aut-Ed-Kd}]
\begin{proof}[Proof of Prop.~\ref{prop:aut-Ed-Kd}]
Let~$F$ be an orthogonal automorphism of~$V = ⨁ E_d ⊗ K_d$.
For any basis of the coefficient spaces~$E_d$,
the restrictions of~$F$ to maps~$K_d → K_{d'}$
are $≻$-preserving in the sense of~\ref{prop:hom-Kd}.
Applying Prop.~\ref{prop:hom-Kd}, we see that
$F$~is of the form given in~\eqref{eq:aut-Ed-Kd}.

Let now $F$~be any linear map defined as in~\eqref{eq:aut-Ed-Kd}.
Writing down the expansion of~$b_{λ} (F(u ⊗ f), F(v ⊗ φ))$,
we see that $F$~is orthogonal if, and only if,
its coefficients~$α_{d',d}$, $β_{d',d}$ and~$γ_{d',d}$
satisfy the relations~\eqref{eq:aut-Ed-Kd-ABC}.
\end{proof}%>>>

\subsection{Action on the diagonal coefficients}%<<<2
Let~$(q_{λ})$ be a quadratic pencil with polar pencil~$(b_{λ})$
isomorphic to~$⨁ E_d ⊗ K_d$ for some coefficient spaces~$E_d$.

Let~$F$ be an automorphism of~$(b_{λ})$ as in Prop.~\ref{prop:aut-Ed-Kd},
and write~$q' = q ∘ F$.
We then have, for all~$u ∈ E_d, f ∈ H_{d-1}, φ ∈ \chk{H_d}$:
\begin{equation}\begin{split}
q_{λ} (F (u ⊗ f))
  &= ∑_{i ≥ d} q(α_{i,d} (u) f) + ∑_{i} q(γ_{i,d} (f u))
  + ∑_{i} b(γ_{i,d} (u), α_{i,d} (u) f^2);\\
q_{λ} (F (u ⊗ φ))
  &= ∑_{i ≤ d} q(β_{i,d} (u) φ),
\end{split}\end{equation}
Solving the IP1S problem means computing the values~$α_{d,d'}$,
$β_{d,d'}$ and~$γ_{d,d'}$ as defined in Prop.~\ref{prop:aut-Ed-Kd}.
In view of the relation~$\transpose{C} · A + \transpose{A} · C = 0$
of this proposition,
we replace the unknowns~$γ_{d,d'}$
by~$γ'_{d,d'} = ∑_i \transpose{α_{i,d}} γ_{i,d'}$;
then $γ_{d,d'}  ∈ \Hom (E_d, E_{d'}) ⊗ \chk{H_{d+d'-1}}$
and the anti-symmetry condition is $\transpose{γ'_{d,d'}} = γ_{d',d}$.

With its polar form known,
the quadratic pencil is determined by its diagonal coefficients.
Write~$ψ_{d}$ for the restriction of~$σ^{-1} ∘ q$ to~$E_d ⊗ H_{d-1}$
and $ω_d$ for its restriction to~$E_d ⊗ \chk{H_d}$;
then $ψ_d$ and~$ω_d$ are pencils of $k$-linear forms and they determine~$q$.
We likewise define~$ψ'_d, ω'_d$ as the diagonal coefficients of~$q'$.

Let~$(ξ_{i,d-i})$ be the basis of~$\chk{H_d}$ dual to
the basis~$(x^i y^{n-i})$ of~$H_D$,
and decompose~$γ_{d,d'}$ in this basis as~$∑_{i,j} γ_{d,d',j} ξ_{j,d+d'-1-j}$,
where~$γ_{d,d',j} ∈ \Hom (E_d, E_{d'})$.
The coefficients of~$γ'_{d,d'}$ are then given by
$γ'_{d,d',j} = ∑_{i,r} \transpose{α_{i,d,r}} · γ_{i,d',r+j}$.

From this we deduce the formula
for the diagonal coefficients~$ψ'_d, ω'_d$ of~$q'$:
\begin{equation}\begin{split}
ψ'_d &= ∑_i ψ_i ∘ α_{i,d} + ∑_i ω_i ∘ γ_{i,d}
  + ∑_{i,j} σ^{-1} (γ'_{d,d,2j+1} - λ γ'_{d,d,2j}) ξ_{j,d-j},\\
ω'_d &= ∑_i ω_i ∘ β_{i,d}.
\end{split}\end{equation}

Up to a replacement of~$σ$ by its inverse~$σ^{-1}$, we obtain the
following proposition.

\begin{prop}\label{prop:ip1s-semilinear}%<<<
The $n$-dimensional IP1S problem for totally irregular pencils over a
field of characteristic two is equivalent to a set of~$O(n^2)$ linear
equations and one semi-linear equation of the form
\begin{equation*}
X = A σ(X) + B,
\end{equation*}
where $A$~is a square matrix and $X, B$~are column matrices of
dimension~$O(n^2)$.
\end{prop}%>>>

\subsection{Solving Frobenius equations}%<<<2

As the matrix~$A$ of Proposition~\ref{prop:ip1s-semilinear}
empirically seems to be of general type,
we give a generic method for this family of \emph{Frobenius equations}.
In the IP1S problem, the base field~$k$ has characteristic~two; we
present here the general case for any (finite) base field.

Let~$k[φ]$ be the non-commutative ring of polynomials in~$φ$,
with the relations~$φ c = σ(c) φ$ for all~$c ∈ k$. This ring is Euclidean.
More precisely, the (left-side) Euclidean algorithm works:
given two elements~$a, b ∈ k[φ]$, there exist elements~$u, v$
such that~$u a + v b = d$ where $d$~is the gcd of~$a$ and~$b$.
Moreover, the only two-sided ideals of~$k[φ]$
are those generated by the powers of~$φ$.

From these two remarks and \cite[Ch.~3, Th.~19]{jacobson1943rings} we get
the following.
\begin{thm}
Let~$φ: k^n → k^n$ be a semi-linear endomorphism. There exists a basis
of~$k^n$ in which the matrix of~$φ$ is the direct sum of cyclic matrices.
% at most one of which is not nilpotent.
\end{thm}
We note that this reduction is also a degenerate case of
the Dieudonné-Manin reduction given by the Newton polygon
for semi-linear endomorphism over the ring of Witt vectors~$W(k)$.

Assume now that $A$~is cyclic and let~$R = k[a]/f(a)$ be the $k$-algebra
generated by~$a$; the equation of Prop.~\eqref{prop:ip1s-semilinear} may
then be written as
\begin{equation}\label{eq:semi-linear-pol}
x = a σ(x) + b,
\end{equation}
where~$a$, $b$ and~$x$ belong to the finite algebra~$k[a]$, and $σ$~is
the absolute Frobenius automorphism. To the primary factorization~$f = ∏
f_i$ of~$f$, there corresponds a factorization~$R = ∏ R_i$ where $R_i$~is
a local algebra. Using Chinese remainders, we may therefore assume that
$R$~is a local algebra and~$f = f_0^d$, where $f_0$~is irreducible.

We first see to the case where~$d = 1$, \emph{i.e.} $R$~is an extension
of the field~$k$. Write~$k_0 = \F_p$ be the field fixed by the
Frobenius~$σ$ and let~$N_0 = N_{R/k_0}$ and~$\Tr_0 = \Tr_{R/k_0}$ be the
norm and trace operators. The equation~\eqref{eq:semi-linear-pol} implies
that
\begin{equation}
x = b' + N_0(a) x, \quad\text{where $b' = b + a σ(b) + … + a σ(a) …
σ^{n-2}(a) σ^{n-1}(b)$.}
\end{equation}
If $N_0(a) ≠ 1$, then this gives as a unique solution~$x =
b'/(1-N_0(a))$. If, on the contrary, $N_0(a) = 1$, then by
Hilbert's theorem~90~[Lang, VI.6.1], there exists~$u ∈ R^{×}$ such
that~$a = σ(u)/u$; then~$x' = ux$ satisfies the equation~$x' = σ(x') +
ub$. This last equation, using the additive form of Hilbert's theorem~90
[Lang, VI.6.3], has a solution if, and only if, $\Tr_0 (ub) = 0$. We
note that both forms of Hilbert's theorem are algorithmic.

In the general case where~$d ≥ 1$, let~$\fr m$ be the maximal ideal
of~$R$. We may use the preceding paragraph to compute a solution~$x_0$ of
the equation modulo~$\fr m$. Moreover, we notice that~$f(x) = x -  a σ(x)
- b$ is a polynomial and that~$f'(x) = 1$; therefore, this polynomial is
separated, and we may use Hensel's lemma to lift the approximate
solution~$x_0$ to a full solution.

%>>>1
\section{Regular quadratic pencils in characteristic two: the regular part}%<<<1
\label{S:quad-reg}
\subsection{Quadratic pencils and extensions of scalars}%<<<2
Let~$k$ be a finite field of characteristic two and $q$~be a regular
quadratic pencil on a $k$-vector space~$V$;
let~$b$ be the polar pencil of~$q$ and~$c$ be its characteristic endomorphism.
We say that a quadratic form \emph{commutes} with a $k$-algebra~$R$ if
its polar form commutes with~$R$ in the sense of
section~\ref{ss:commute}.
In this way, the pencil~$q$ determines a quadratic form~$q_{∞}$
commuting with the local algebra~$R = k[c]$.
In particular, the polar form~$b_{∞}$ is an alternating bilinear form
commuting with~$R$.

The following proposition is an intrinsic, and more general, form
of the result already known for cyclic pencils~\cite[Prop.~5]{MPG2013}.

\begin{prop} \label{prop:trace-quad}%<<<
Let~$K$ be a finite separable extension of~$k$ and~$V$ be a $K$-vector
space. For any $k$-quadratic form~$q: V ⊗ V → k$ commuting with~$K$,
there exists a unique $K$-quadratic form~$q_K: V ⊗ V → K$ such that~$q =
\Tr_{K/k} \, ∘ \, q_K$.
\end{prop}

\begin{proof}
Let~$b$ be the polar form of~$q$. By~\ref{prop:trace-form},
there exists~$b_K ∈ K$
such that~$b(x,y) = \Tr_{K/k} (b_K xy)$; in particular, since $b$~is
alternating, for all~$x$ we have~$\Tr_{K/k} (b_K x^2) = 0$ and
therefore~$b_K = 0$. Therefore, $q$~is a semi-linear form; since the
trace map is non-degenerate, there exists~$q_K$ such that~$q(x) =
\Tr_{K/k} (q_K x^2)$.

As for Prop.~\ref{prop:trace-form}, the $n$-dimensional case directly follows by
taking coordinates. Here all diagonal entries correspond to quadratic
forms, and all others to bilinear forms; all of these commute with~$K$.
\end{proof}%>>>

\subsection{Alternating forms commuting with a local algebra}%<<<2
\label{SS:alt-pencil}

Using Prop.~\ref{prop:trace-quad},
we assume that~$K = k$ and write~$R = k[π] / π^ℓ$.
We equip this algebra with the Frobenius and Verschiebung automorphisms
defined for~$x ∈ k$ by~$σ(x) = x^2$ and~$V(x) = x$,
and by~$σ(π) = π$ and~$V(π) = π^2$.
We note that, for all~$x ∈ R$, we have~$V σ(x) = σ V(x) = x^2$;
moreover, $R$~decomposes as~$R = V(R) ⊕ π V(R)$.

We recall that the map~$τ: R → k$ given by the coefficient of~$π^{ℓ-1}$
produces the regular $k$-bilinear form~$τ(x y)$ on~$R$.
We call a $R$-bilinear form~$b$ \emph{$τ$-alternating} if $τ ∘ b$~is
alternating, and say that an element~$a$ of~$R$ is $τ$-alternating if the
bilinear form~$a x y$ is.
The space~$R^{τ}$ of $τ$-alternating elements of~$R$ is linearly spanned
by the~$π^{ℓ-2i}$ for~$0 ≤ i ≤ ℓ/2$.
A $R$-bilinear form~$b$ with matrix~$B$ is $τ$-alternating if, and only if,
$B$~is anti-symmetric and all its diagonal coefficients are $τ$-alternating.

Let~$(b_0, b_{∞})$ be a pencil of alternating $k$-bilinear forms on~$V$,
with characteristic endomorphism~$π$. The $R$-bilinear form~$b_R = b_{∞,
R}$ associated to~$b_{∞}$ by~\ref{prop:trace-local} is then
$τ$-alternating. Moreover, the $R$-bilinear form associated to~$b_{0}$
is~$b_{0, R} = π b_R$; since $b_{0}$~is alternating, $π b_R$ is again
alternating. Therefore, since $b_R$~and~$π b_R$ are $τ$-alternating,
$b_R$~is an alternating form.

Prop.~\ref{prop:diag-bigblock} applies to $τ$-alternating forms and shows
that they are an orthogonal sum of regular $τ$-alternating forms on free
modules over quotient rings of~$R$. We therefore assume that $V$~is free
as a $R$-module.

\paragraph{Classification of $τ$-alternating forms.}
% Classification by the norm
Although this is not directly useful for the IP1S problem,
$τ$-alternating forms have an elegant classification up to congruence.
For any bilinear form~$b$ on a free $R$-module~$M$, we define the
\emph{norm} of~$b$ as the ideal~$\fr{n} b$ of~$R$
generated by the elements~$b(x, x)$ for~$x ∈ M$.

\begin{prop}\label{prop:eqv-norm}%<<<
Two non-degenerate, $τ$-alternating forms are equivalent if and only
if they have the same norm.
\end{prop}

\begin{proof}
We show that the bilinear form~$b$ is equivalent to either the form with
identity matrix, if $\fr n b = R$; or the orthogonal direct sum
\[ N_a = \mat{a & 1\\1&0} ⟂ \mat{0 & 1\\1&0} ⟂ … ⟂ \mat {0 & 1\\ 1 & 0}, \]
where $a$~is a generator of~$\fr n b$, if $\fr n b ⊂ π R$.
We note that, when the dimension of~$V$ is odd, the identity form is
congruent to the form~$N_1$, since the transformation~$\mat{1&1\\0&1}$
maps the identity matrix to the form~$\mat{1&1\\1&0}$.

We write~$[u]$ for the one-dimensional form with coefficient~$u$
and~$H_{v, w}$ for the form with matrix~$\mat{v&1\\1&w}$.
We also write~$H = H_{0,0}$.

By~\cite[§2]{milnor2}, the $k$-bilinear form~$b_k = b ⊗_R k$ has a
decomposition~$b_k ≃ [d_1] ⟂ … ⟂ [d_r] ⟂ H^s$, where all~$d_i$ are
non-zero since $b$~is non-degenerate. We distinguish two cases.

\subparagraph{Case 1: $\fr n b = R$.}
We first show that the reduced form~$b_k$ is isomorphic, over~$k$, to the
identity form. Since $\fr n b = R$, the list of~$d_i$ in the above
decomposition is not empty. Moreover, since the field $k$~is perfect, all
elements~$d_i$ are squares in~$k$, so that up to a coordinate change we
may assume that~$d_i = 1$. Finally, we note that the
matrix~$\mat{1&1&0\\1&0&-1\\1&1&-1}$ is an isomorphism between the
forms~$[1] ⟂ [1] ⟂ [-1]$ and~$[1] ⟂ H$, so that if $r ≥ 1$ we may cancel
all the direct factors~$H$ and in this case $b_k$~is isomorphic to the
bilinear form with identity matrix.

By~\cite[Corollary 3.4]{baeza1978quadratic},
the diagonalization of~$b_k$ lifts to a diagonalization
$b ≃ [a_1] ⟂ … ⟂ [a_r]$ over~$R$.
Since $b$~is $τ$-alternating, all coefficients~$a_i$ are $τ$-alternating.
If the length of~$R$ is odd, then the set of $τ$-alternating elements of~$R$
is~$π V(R)$, which contradicts the regularity of~$b$.
Therefore, the length of~$R$ is even, which means that $a_i ∈ V(R)$ and they are
therefore squares in~$R$.
From this we deduce that $b$~is isomorphic to the identity bilinear form.

\subparagraph{Case 2: $\fr n b ⊂ π R$.}
This means that the form~$b_k$ is alternating, so that~$b_k ≃ H^s$.
By~\cite[Corollary 3.4]{baeza1978quadratic}, this decomposition again lifts to a
decomposition $b ≃ H_{u_1, v_1} ⟂ … ⟂ H_{u_s, v_s}$, where all
coefficients~$u_i, v_i$ are $τ$-alternating.

Let $b = H_{u,v}$ be $τ$-alternating and regular; up to a swap of~$u, v$, we
may write it as~$H_{u, u a^2}$ for some~$a ∈ R$. Since $b$~is
regular, $1 + a u ∈ R^{×}$. Therefore, the coordinate
change $P = \mat{1+a u & a/(1+a u) \\ u & 1/(1+a u)}$ transforms the
bilinear form~$b$ to~$H_{u, 0}$.

Finally, we note that the bilinear form~$b = H_{u, 0} ⟂ H_{u a^2 u, 0}$ is
isomorphic to~$H_{v, 0} ⟂ H_{0,0}$ via the coordinate change
% \begin{equation}
$\mat{1 & 0 & a & 0\\0&1&0&0\\0&0&1&0\\au & a & 0 & 1}$.
% \end{equation}
\end{proof}%>>>

\subsection{Reduction of local quadratic pencils}%<<<2

We first give an explicit description of quadratic forms commuting
with~$R$.

\begin{lem}\label{lem:gamma-polar}%<<<
Let~$γ: R → R$ be the application defined, for~$x = V(y) + π V(z)$,
by~$γ(x) = π V(yz)$. Then, for all~$u, x ∈ R$, we have
\begin{enumerate}
\item $γ(ux) = γ(u) x^2 + u^2 γ(x)$;
\item $γ(u+x) = γ(u) + γ(x) + ux + V(α)$ for some~$α ∈ R$.
\item For all~$c ∈ R$, $τ(c γ (x))$~is a $k$-quadratic form on~$R$
with polar~$τ(c x y)$.
\end{enumerate}
\end{lem}

\begin{proof}
(i)~Write~$x = V(y) + π V(z)$ and~$u = V(v) + π V(w)$. Noticing
that~$σ(x) = y^2 + π z^2$, we then have
\begin{equation}\label{eq:gamma-prod}
\begin{split}
γ(ux) &= π V\pa{(vy + π wz)(wy+vz)}\\
 &= π V\pa{vw (y^2 + π z^2)} + π V \pa{(v^2+πw^2) yz}\\
 &= γ(u) V(σx) \;+\; V(σu) γ(x) \quad = γ(u) x^2 + u^2 γ(x).
\end{split}
\end{equation}

(ii) follows from
$γ(u+x) - γ(u) - γ(x) = π V(vz + wy) = ux - V(vy+πwz)$.

(iii)~is a direct consequence of~(i) and~(ii).
\end{proof}%>>>
\begin{prop}\label{prop:quad-tau}%<<<
Let~$M = ⨁ R_{m_i}$ be a $R_ℓ$-module of finite length.
Let~$q$ be a $k$-quadratic form on~$M$ commuting with~$R$, $b$~its
polar form, and $b_R$ be the unique $R$-quadratic form such that~$b = τ ∘
b_R$; write~$b_{i,j}$ for the coefficients of~$b_R$.
Then there exists a $σ$-linear form~$a: M → R$ such that, for all~$x =
(x_i) ∈ M$,
\begin{equation}
q(x) \;=\; τ\pa{a(x) + ∑_{i} b_{i,i} γ(x_i) + ∑_{i < j} b_{i,j} x_i x_j}.
\end{equation}
\end{prop}

\begin{proof}
We see by Lemma~\ref{lem:gamma-polar} and by linearity that
% $τ(c γ(x))$~is a quadratic form on~$R$ with polar~$τ(c\, xy)$.
% It follows by linearity that
$q' = τ(∑ b_{i,i} γ(x_i) + ∑ b_{i,j} x_i x_j)$~is a $k$-quadratic form on~$M$,
which has the same polar as~$q$,
so that the difference~$q - q'$ is a $σ$-linear form.
\end{proof}%>>>

Since the bilinear form~$τ(x y)$ is regular on~$R$, the $k$-linear
map~$V: R → R$ has an adjoint~$θ$, such that~$τ(x\, V(y)) = τ(θ(x)\, y)$.
This map is defined by~$θ(π^{ℓ-1-2i}) = π^{ℓ-1-i}$ and~$θ(π^{ℓ-2i}) = 0$.

When computing the effect of a $R$-linear change of variable on a
$k$-quadratic form commuting with~$R$, we obtain the following result.

\begin{prop}\label{prop:quad-changevar}%<<<
Let~$u: M' → M$ be a $R$-linear map between two $R$-modules of finite length.
Let~$q$ be a $k$-quadratic form on~$M$, commuting with~$R$, and~$q' = q ∘ u$.
Let~$b$ be the polar form of~$q$ and~$a = (a_i)$ be the $σ$-linear form defined
as in Prop.~\ref{prop:quad-tau}; then the corresponding values~$b', a' =
(a'_i)$
for~$q'$ are
\begin{equation*}
b' = b ∘ u, \quad
a'_i \;=\; ∑_r a_r σ(u_{r,i})
  + θ \pa{ ∑_r b_{r,r} γ(u_{r,i}) + ∑_{r < s} b_{r,s} u_{r,i} u_{s,i} }.
\end{equation*}
\end{prop}%>>>

The \emph{hyperbolic plane} is the module $R^2$, equipped with the
$k$-quadratic form~$q(x) = τ(x_1 x_2)$; this form commutes with~$R$. The
corresponding $τ$-alternating bilinear form~$b_R$ has the
matrix~$\mat{0&1\\1&0}$. A quadratic form is \emph{hyperbolic} if it is
isomorphic to the orthogonal sum of copies of the hyperbolic plane.
We say that a quadratic form~$q$ \emph{reduces} to a form~$q'$ if $q$~is
isomorphic to the direct sum of~$q'$ and a hyperbolic space.

\begin{prop}\label{prop:witt-four}%<<<
Let~$M$ be a free $R$-module and $q$~be a $k$-quadratic form on~$M$,
commuting with~$R$.
If the polar form of~$q$~is hyperbolic, then $q$~reduces to a quadratic
form of dimension at most two.
\end{prop}

\begin{proof}
Since the polar form of~$q$ is hyperbolic,
there exist some coefficients~$a_1, …, a_{m}; a'_1, …, a'_{m} ∈ R$
such that $q$~is isomorphic to the form
$q' = [a_1, …, a'_m]$ defined on~$R^{2m}$ by
\begin{equation}
q' (x_1, …, x_m, x'_1, …, x'_{m}) \;=\;
  τ\pa{∑_{i ≤ m} a_i σ(x_i) + a'_i σ(x'_i) + γ(x_i x'_i)}.
%   τ(∑_{i ≤ 2m} a_i σ(x_i) + ∑_{i ≤ m} γ(x_{i} x_{i+m})).
\end{equation}
To prove the proposition, we show that
any form~$q = [a_1, a_2; a'_1, a'_2]$ is isomorphic to
a form~$[b, 0; b', 0]$ for some~$b, b' ∈ R$.
The polar of~$q$ is the hyperbolic form on~$R^4$,
and its automorphism group
% \begin{equation}
% H_4 = \mat{0&1&0&0\\-1&0&0&0\\0&0&0&1\\0&0&-1&0}.
% \end{equation}
% The automorphism group of~$H_4$
contains the following transformations,
with the corresponding effect on the values~$(a_i)$:
\begin{equation}\begin{array}{ll}
\mat{ & & 1 &\\ &&&1\\ 1&&&\\ &1&&} &
\begin{cases}a_1 \leftrightarrow a'_1\\
  a_2 \leftrightarrow a'_2\\\end{cases} \\
\mat{M & 0\\ 0 & \transpose{M}^{-1}}, M ∈ \GL_2(R) &
\begin{cases} (a_1, a_2) ← (a_1, a_2) · σ(M)\\
(a'_1, a'_2) ← (a'_1, a'_2) · σ(\transpose{M}^{-1}) \end{cases}\\
\mat{1&&u\\&1&&v\\ &&1&\\ &&&1} &
\begin{cases} a'_1 ← a'_1 + a_1 \, σ(u) \,+\, θ(u) \\
a'_2 ← a'_2 + a_2 \, σ(u) \,+\, θ(v)\end{cases}\\
\mat{1&&&w\\&1&w&\\ &&1&\\ &&&1} &
\begin{cases} a'_1 ← a'_1 + a_2 \, σ(w) \\
a'_2 ← a'_2 + a_1 \, σ(w) \end{cases}
\end{array}\end{equation}
Using the first transformation, we may assume that
the ideal~$(a_1, a_2)$ of~$R$ contains~$(a'_1, a'_2)$.
Using the second one, we may assume that $a_2 = 0$,
so that $a_1$~divides $a'_1$ and~$a'_2$.
Finally, using the last transformation, one may have~$a'_2 = 0$.
\end{proof}%>>>

Using Prop.~\ref{prop:witt-four} on both quadratic forms of a quadratic
pencil, we see that any quadratic pencil reduces to a pencil of dimension
at most four.
Therefore, to compute a linear equivalence between any two quadratic
pencils, we may assume that the dimension of the ambient $R$-module is at
most four.

\begin{prop}\label{prop:ip1s-bin-polynomial}%<<<
Let~$k$ be a finite field of characteristic two.
There exists a polynomial algorithm computing a linear isomorphism between
two equivalent quadratic pencils over~$k$.
\end{prop}

\begin{proof}
Let~$(q_{λ})$, $(q'_{λ})$ be two equivalent quadratic pencils.
By subsection~\ref{SS:alt-pencil}, we may assume that
the associated polar pencils~$(b_{λ})$, $(b'_{λ})$
define an hyperbolic alternating form on~$R^{2n}$.

Using Prop.~\ref{prop:witt-four} for the form~$q_{∞}$,
we may assume that $q_{∞}$~is of the form~$[a, 0, …, 0; a', 0, …, 0]$.
Using Prop.~\ref{prop:witt-four} again for the restriction of~$q_{0}$
to the coordinates with index~$(2, …,  n; n+2, …, 2n)$ of~$R^{2n}$,
we may further assume that
$q_{0}$~is of the form~$[b, c, 0, …, 0; b', c', 0, …, 0]$.
This shows that the pencil~$(q_{λ})$ is isomorphic to
the direct sum of a pencil of dimension at most~$4$
and a hyperbolic pencil.
The same applies to~$(q'_{λ})$.
Since both hyperbolic parts are isomorphic,
we only need to compute the isomorphism for pencils of dimension~$≤ 4$.

Let~$q$ be the quadratic form defined by
\begin{equation}
q(x_1, …, x_4) = τ (∑_{i} a_i σ(x_i) + x_1 x_3 + x_2 x_4)
\end{equation}
and likewise, let~$q'$ be a quadratic form linearly equivalent to~$q$,
with the same hyperbolic polar form~$b$,
and with diagonal coefficients~$a'_i$.

An isomorphism~$u: q → q'$ is given by coefficients~$u_{i,j}$
satisfying the equations of Prop.~\ref{prop:quad-changevar}.
Writing~$u = V(v) + π V(w)$, and likewise for~$u'$ and~$u_{i,j}$, we have
\begin{equation}
σ (u) = v^2 + π w^2, \qquad
θ(a u u') = (v v' + π w w') θ(a) + (v w' + w v') θ(π a),
\quad\text{and}\quad
θ(a γ (u)) = v\, w\, θ(a π),
\end{equation}
so that the equations on the 16 variables~$u_{i,j}$ are equivalent to
the following polynomial equations
in the 32 variables~$v_{i,j}$ and~$w_{i,j}$:
\begin{align}\label{eq:ip1s-polynomial}
σ (b_{i,j}) &= ∑_{r,s} σ (b_{r,s})\: (v_{r,i}^2 + π w_{r,i}^2)\:
  (v_{s,j}^2 + π v_{s,j}^2),\\
a'_{i} &= ∑_{r} a_{r} (v_{r,i}^2 + π w_{r,i}^2)
  + ∑_{r < s} (v_{r,i} v_{s,i} + π w_{r,i} w_{s,i}) θ (b_{r,s})
  + (v_{r,i} w_{s,i} + w_{r,i} v_{s,i}) θ (π b_{r,s}).
\end{align}
This shows that the IP1S problem is equivalent to
a bounded number of polynomial equations
in a bounded number of variables in the ring~$R = R_{ℓ} = K[π] /π^{ℓ}$
(more precisely, it is enough
to determine~$v_{i,j}$ up to precision~$\floor{ℓ/2}$
and~$w_{i,j}$ up to precision $\floor{(ℓ-1)/2}$).
Since $R$~is a discrete valuation ring,
linear equations are solvable in~$R$
in the sense of~\cite[4.1.5]{adams1994grobner};
therefore, it is possible to compute a Gröbner basis of
the ideal generated by the equations~\ref{eq:ip1s-polynomial}
in~$R[v_{i,j}, w_{i,j}]$.
Computing this basis is possible with a number of ring operations
polynomial in the degree of the equations
and doubly exponential in the number of variables~\cite{dube1990grobner}.
However, in our case there are only 32~variables
and the equations are homogeneous of degree~two.
Therefore, the computation of the Gröbner basis
requires a bounded number of computations in the ring~$R$;
each of these computations requires a polynomial in~$n$
number of computations in the base field~$k$.
\end{proof}%>>>

\section{Computation of the second secret for IP2S}%<<<1
\label{S:IP2S}

\subsection{Reduction to the regular case}
Two families of polynomials~$(a_1,…,a_m)$ and~$(b_1,…,b_m)$ are
\emph{isomorphic with two secrets} if there exist bijective linear
transformations~$s$ of the $n$~variables and~$t$ of the $m$~polynomials
such that $t ∘ \bm{a} ∘ s = \bm{b}$. Assume that $m = 2$. Then the second
secret~$t$ is a homography in two variables, which we write~$γ ∈
\GL_2(k)$.

\begin{prop}
\def\reg{'}
Let~$\bm{a}$, $\bm{b}$ be two pencils and $\bm{a}\reg$ and~$\bm{b}\reg$ be
their regular parts. For any homography~$t$, $t ∘ \bm{a}$~is isomorphic
(in the IP1S sense) to~$\bm{b}$ if, and only if, $t ∘ \bm{a}\reg$~is
isomorphic to~$\bm{b}\reg$.
\end{prop}

\begin{proof}
The minimal index of the pencil~$\bm{b}$ is the minimal degree of an
isotropic vector~$e_0 + … +λ^h e_h$ for~$b_{λ}$; such a vector may be
written in homogeneous form in~$(λ:μ)$ as~$e(λ:μ) = ∑ λ^i μ^{h-i} e_i$,
which is isotropic for the quadratic form~$b(λ:μ) = μ b_0 + λ b_{∞}$. Now
let~$γ = \smat{a&b\\c&d} ∈ \GL_2(k)$ be a homography. Then the
vector~$e^{γ}$ defined by~$e^{γ}(λ:μ) = e(aλ+bμ:cλ+dμ)$~is isotropic
for~$b_{γ(λ)}$ iff $e$~is isotropic for~$b$. This proves that the pencils
$(b_{γ(λ)})$ and~$(b_{λ})$ have the same minimal index. Therefore, all
their Kronecker blocks coincide.
\end{proof}

% \begin{prop}
% Let~$K_{λ}$ be a Kronecker block of dimension~$2h+1$ as in
% Prop.~\ref{prop:kronecker1}. Then for all~$γ ∈ \GL_2(k)$, the pencils
% with matrix~$K_{λ}$ and~$K_{γ(λ)}$ are isomorphic in the IP1S sense.
% \end{prop}
% 
% \begin{proof}
% Let~$(e_0,…,e_h; f_1,…,f_h)$ be the basis in which the pencil~$(b_{λ})$
% has the Kronecker-block matrix. Let~$e(x) = ∑ x^i e_i$ and~$f(y) = ∑
% y^{i-1} e_i$. The pencil~$(b_{λ})$ is then defined by the relation
% \begin{equation}\label{eq:def-b-intr}
% b_{λ} (e(x), f(y)) \;=\; (x+λ) F(xy), \quad F(t) = 1 + … + t^{h-1}.
% \end{equation}
% For any homography~$γ = \smat{a&b\\c&d} ∈ \GL_2(k)$, define a
% basis~$(e_i^{γ}, f_i^{γ})$ by the relations
% \begin{equation}\label{eq:def-ei-gamma}
% ∑ (a λ + b μ)^i (c λ + d μ)^{h-i} e_i = ∑ λ^i μ^{h-i} e_i^{γ}.
% \end{equation}
% We then find that in the basis 
% 
% ; then the only relation is~$b_{λ} · ∑ λ^i
% e_i = 0$, which we write as a homogeneous polynomial
% in~$(λ:μ)$ as $(λ b_{∞} + μ b_{0}) (∑ λ^i μ^{h-i} e_i) = 0$.
% Let~$γ = \smat{a&b\\c&d}$ and define a basis~$(e_i^{γ}, f_i^{γ})$ by
% Then in the basis
% \end{proof}

\subsection{IP2S in the regular case}

Let~$(a_{λ})$ and~$(b_{λ})$ be two \emph{regular} pencils of bilinear
forms such that $a_{γ(λ)}$~is isomorphic, in the IP1S sense, to~$b_{λ}$.
Then the homography~$γ$ maps the characteristic polynomial~$f(λ:μ)
= \det (a_{λ:μ})$ to~$g(λ:μ) = \det (b_{λ:μ})$. In particular, it maps the
prime factors of~$f$ to those of~$g$, respecting both their degree and
their exponent as a factor of the characteristic polynomial.

Let~$S_{d,e}$~and~$T_{d,e}$ be the set of factors of degree~$d$ and
exponent~$e$ of the polynomials~$f$ and~$g$. Then any homography~$γ$
mapping all the elements of~$S_{d,e}$ to~$T_{d,e}$ for each pair~$(d,e)$
is a possible second secret in the IP2S problem. We compute the
intersection for~$(d,e)$ of the set~$Γ_{d,e}$ of homographies mapping the
prime polynomials of~$S_{d,e}$ to~$T_{d,e}$. In most cases, the first
set~$Γ_{d,e}$ already contains only one candidate, which is therefore the
second secret~$γ$. The discussion depends on the degree~$d$ of the
polynomials. We note that the sum of the size of the sets~$S_{d,e}$ is
the number of variables~$n$; therefore, we may use the worst-case
estimate~$\card{S_{d,e}} = O(n)$ for each~$(d,e)$.

We shall use the following classic results.
\begin{prop}\label{prop:homography}
\begin{enumerate}
\item Let~$(x_1, x_2, x_3)$ and~$(y_1, y_2, y_3)$ be two (ordered)
triples of distinct points of~$ℙ^1(k)$. There exists a unique
homography~$γ ∈ \mathrm{PGL}_2(k)$ such that~$γ(x_i) = y_i$.
\item Let~$(x_1, x_2, x_3, x_4)$ and~$(y_1, y_2, y_3, y_4)$ be two
(ordered) quadruplets of distinct points. They are homographic iff they
have the same cross-ratio~$B(x) = B(y)$, where
\begin{equation}
B(x) = \frac{(x_1-x_3)(x_2-x_4)}{(x_1-x_4)(x_2-x_3)}.
\end{equation}
\item Let~$\acco{x_1, x_2, x_3, x_4}$ and~$\acco{y_1, y_2, y_3, y_4}$ be
two (unordered) sets of four points. They are homographic iff they have
the same $j$-invariant~$j(x) = j(y)$, where
\begin{equation}\label{eq:j-invariant}
j(x) = \frac{(B(x)^2-B(x)+1)^3}{B(x)^2(1-B(x))^2}.
\end{equation}
\item Let~$u(x) = ∑ u_i x^i$ and~$v(x)$ be two monic polynomials
of degree four. They are homographic iff they have the same
$j$-invariant, where $j(u)$~is a rational function of degree~six in the
coefficients of~$u$.
\end{enumerate}
\end{prop}

We note that the formula for the $j$-invariant given
in~\eqref{eq:j-invariant} is, up to a constant factor, the formula for
the $j$-invariant of an elliptic curve. Namely, two elliptic curves with
equations~$y^2 = f(x)$ and~$y^2 = g(x)$, where $f, g$ are separable
polynomials of degree~$≤ 4$, are isomorphic iff the polynomials~$f$
and~$g$ are homographic.

\bigbreak
We now explain how we compute the set~$Γ_{d,e}$ for each pair~$(d,e)$.

\paragraph{Case~$d = 1$.}
If $\card{S_{1,e}} ≥ 3$, then we may immediately recover the
homography~$γ$: namely, fix a triple~$(x_1, x_2, x_3)$ in~$S_{1,e}$, and
iterate over the triples in~$T_{1,e}$. For each such triple, there exists
a unique homography~$γ$ such that~$γ(x_i) = y_i$. This homography belongs
to~$Γ_{1,e}$ iff the images of all the other points of~$S_{1,e}$ belong
to~$T_{1,e}$. Since there are~$3!\binom{\card{S_{1,e}}}{3} = O(n^3)$
triples~$(y_i)$, this computation requires~$O(n^3)$ field operations.

If $1 ≤ \card{S_{1,e}} ≤ 2$, then $Γ_{1,e}$~may be explicitly computed as
the union of the set of homographies mapping the elements of~$S_{1,e}$ to
those of~$T_{1,e}$ for all permutations of~$T_{1,e}$.

\paragraph{Case~$d = 2$.}
Assume $\card{S_{2,e}} ≥ 2$. Let~$u_1, u_2 ∈ S_{2,e}$ and~$v_1, v_2 ∈
T_{2,e}$ be monic polynomials of degree~two. Any homography between the
sets~$\acco{u_1, u_2}$ and~$\acco{v_1, v_2}$ will map~$u_1 u_2$ to~$v_1
v_2$. By Prop.~\ref{prop:homography}(iv), there exists at most a bounded
number of such homographies. Since there are~$\binom{\card{S_{2,e}}}{2} =
O(n^2)$ pairs~$(v_1, v_2)$, this requires~$O(n^2)$ field operations.

If~$\card{S_{2,e}} = 1$, then $Γ_{2,e}$~is the set of all homographies
mapping the unique element of~$S_{2,e}$ to the unique element
of~$T_{2,e}$.

\paragraph{Case~$d = 3$.}
Fix an element~$u ∈ S_{3,e}$. For all~$v ∈ T_{3,e}$, there exist at
most~$3! = 6$ homographies~$γ$ mapping~$u$ to~$v$. Each candidate belongs
to~$Γ_{3,e}$ iff it maps all other elements of~$S_{3,e}$ to elements
of~$T_{3,e}$. There are~$\card{S_{3,e}} = O(n)$ candidates~$u$ and
therefore~$O(n)$ candidate homographies~$γ$.

\paragraph{Case~$d = 4$.}
Fix an element~$u ∈ S_{4,e}$. The candidates as homographic images of~$u$
in~$T_{4,e}$ are the~$v$ such that~$j(v) = j(u)$. Each candidate
polynomial~$v$ gives at most $4! = 24$~candidates homographies~$γ$. This
allows to compute~$Γ_{4,e}$ in~$O(n)$ field operations.

\paragraph{Case~$d ≥ 5$.} The naïve method is to differentiate~$(d-4)$
times the elements of~$S_{d,e}$ to reduce to the case where~$d = 4$.
However, as this uses only the five leading coefficients, if the
polynomials are specially chosen we may find too many homographies; for
example, although the polynomials~$x^d-1$ and~$x^d$ are not homographic,
all their derivatives are. Instead, we first compose all the elements
of~$S_{d,e}$ and~$T_{d,e}$ by a known, randomly chosen homography~$r$. In
general, for any two non-homographic elements~$u_1, u_2 ∈ S_{d,e}$, the
derivatives~$(∂/∂x)^4\, (u_i ∘ r)$ are non-homographic. In the improbable
case where they are homographic, we only need to change the random
homography~$r$. In this way, we may compute the set~$Γ_{d,e}$ in at
most~$O(n)$ field operations.

\paragraph{Computing the hidden homography.}

The hidden homography~$γ$ lies in the intersection of all sets~$Γ_{d,e}$.
As each one of these sets is likely to be extremely small or even reduced
to~$\acco{γ}$, we compute them in increasing order of assumed complexity.
We use the above estimates: for each~$(d,e)$, we use the assumed complexity
\begin{equation}
C_{d,e} = \begin{cases}
\card{S_{d,e}}^3,& d = 1;\\
\card{S_{d,e}}^2,& d = 2;\\
\card{S_{d,e}},& d≥ 3,
\end{cases}
\end{equation}
and sort the pairs~$(d,e)$ by increasing values of~$C_{d,e}$. We finally
find a bounded number of candidate homographies using no more
than~$O(n^3)$ operations in~$k$.

% Conclusion
\section*{Conclusion}

In this paper, we show that we can solve in polynomial-time the IP
problem with two quadratic forms in a finite field of odd characteristic.
The obvious questions are whether it is possible to generalize this to
fields of characteristic two and to more than two equations.

The case of a binary base field is very important for cryptographic
applications. The cyclic case was solved in~\cite{MPG2013}. To solve
the general case, at least two roadblocks remain:
% the Kronecker form
% for the singular part of a bilinear pencil does not apply, as the proof
% of Prop.~\ref{prop:kronecker1} above requires that~$2 ≠ 0$;
quadratic
forms over a local algebra behave differently~\cite[§93]{omeara};
finally, extending from bilinear to quadratic forms requires a study of
the action of a symplectic group on the diagonal coefficients, and this
group becomes quite impractical in the non-cyclic case.

On the other hand, studying the general problem with $m ≥ 3$~quadratic
equations departs from the classic results about pencils of quadratic
forms; therefore, fewer tools are available. Even in the regular
case, our work heavily uses the factorization of the characteristic
polynomial. An analogous strategy for~$m ≥ 3$ would require a detailed
geometric study of the hypersurface defined by this characteristic
polynomial.

%>>>1
\bibliographystyle{plain}
\bibliography{biblio}

\begin{thebibliography}{10}

\bibitem{adams1994grobner}
William~W Adams and Philippe Loustaunau.
\newblock {\em {An introduction to {G}r{\"o}bner bases}}, volume~3.
\newblock American Mathematical Society Providence, 1994.

\bibitem{DBLP:conf/stacs/AgrawalS06}
Manindra Agrawal and Nitin Saxena.
\newblock Equivalence of {F}-algebras and cubic forms.
\newblock In Bruno Durand and Wolfgang Thomas, editors, {\em STACS}, volume
  3884 of {\em Lecture Notes in Computer Science}, pages 115--126. Springer,
  2006.

\bibitem{baeza1978quadratic}
Ricardo Baeza.
\newblock {\em Quadratic forms over semilocal rings}.
\newblock Springer, 1978.

\bibitem{beelen1988improved}
Th~Beelen and Paul Van~Dooren.
\newblock An improved algorithm for the computation of {Kronecker}'s canonical
  form of a singular pencil.
\newblock {\em Linear Algebra and its Applications}, 105:9--65, 1988.

\bibitem{DBLP:journals/corr/BerthomieuFP13}
J{\'e}r{\'e}my Berthomieu, Jean-Charles Faug{\`e}re, and Ludovic Perret.
\newblock Polynomial-time algorithms for quadratic isomorphism of polynomials.
\newblock {\em CoRR}, abs/1307.4974, 2013.

\bibitem{DBLP:conf/pkc/BouillaguetFFP11}
Charles Bouillaguet, Jean-Charles Faug{\`e}re, Pierre-Alain Fouque, and Ludovic
  Perret.
\newblock {Practical Cryptanalysis of the Identification Scheme Based on the
  Isomorphism of Polynomial with One Secret Problem}.
\newblock In Dario Catalano, Nelly Fazio, Rosario Gennaro, and Antonio
  Nicolosi, editors, {\em Public Key Cryptography}, volume 6571 of {\em Lecture
  Notes in Computer Science}, pages 473--493. Springer, 2011.

\bibitem{DBLP:conf/asiacrypt/BouillaguetFM11}
Charles Bouillaguet, Pierre-Alain Fouque, and Gilles Macario-Rat.
\newblock Practical key-recovery for all possible parameters of {SFLASH}.
\newblock In Dong~Hoon Lee and Xiaoyun Wang, editors, {\em ASIACRYPT}, volume
  7073 of {\em Lecture Notes in Computer Science}, pages 667--685. Springer,
  2011.

\bibitem{DBLP:conf/eurocrypt/BouillaguetFV13}
Charles Bouillaguet, Pierre-Alain Fouque, and Amandine V{\'e}ber.
\newblock {Graph-Theoretic Algorithms for the "Isomorphism of Polynomials"
  Problem}.
\newblock In Thomas Johansson and Phong~Q. Nguyen, editors, {\em EUROCRYPT},
  volume 7881 of {\em Lecture Notes in Computer Science}, pages 211--227.
  Springer, 2013.

\bibitem{dube1990grobner}
Thomas~W Dub{\'e}.
\newblock The structure of polynomial ideals and {G}r{\"o}bner bases.
\newblock {\em SIAM Journal on Computing}, 19(4):750--773, 1990.

\bibitem{DBLP:conf/eurocrypt/FaugereP06}
Jean-Charles Faug{\`e}re and Ludovic Perret.
\newblock {Polynomial Equivalence Problems: Algorithmic and Theoretical
  Aspects}.
\newblock In Serge Vaudenay, editor, {\em EUROCRYPT}, volume 4004 of {\em
  Lecture Notes in Computer Science}, pages 30--47. Springer, 2006.

\bibitem{Gantmacher}
F.R. Gantmacher.
\newblock {\em Theory of Matrices}.
\newblock Chelsea, New York, 1960.

\bibitem{gauss}
Carl~Friedrich Gauss.
\newblock {\em Disquisitiones Arithmaticae}.
\newblock Gerhard Fleischer, 1801.

\bibitem{DBLP:journals/jacm/GoldreichMW91}
Oded Goldreich, Silvio Micali, and Avi Wigderson.
\newblock Proofs that yield nothing but their validity for all languages in
  {NP} have zero-knowledge proof systems.
\newblock {\em J. ACM}, 38(3):691--729, 1991.

\bibitem{jacobson1943rings}
Nathan Jacobson.
\newblock {\em The theory of rings}.
\newblock Number~2. American Mathematical Soc., 1943.

\bibitem{kaltoffen11compute}
Erich~L. Kaltofen and Arne Storjohann.
\newblock The complexity of computational problems in exact linear algebra.
\newblock In Björn Engquist, editor, {\em Encyclopedia of Applied and
  Computational Mathematics}. Springer, 2011.

\bibitem{lang-algebra}
Serge Lang.
\newblock {\em Algebra}.
\newblock Springer-Verlag.

\bibitem{lidl1997finite}
R.~Lidl and H.~Niederreiter.
\newblock {\em Finite Fields}.
\newblock Number vol.~20,ptie.~1 in Encyclopedia of Mathematics and its
  Applications. Cambridge University Press, 1997.

\bibitem{MPG2013}
Gilles Macario-Rat, J{\'e}r{\^o}me Pl{\^u}t, and Henri Gilbert.
\newblock {New Insight into the Isomorphism of Polynomial Problem IP1S and Its
  Use in Cryptography}.
\newblock In Kazue Sako and Palash Sarkar, editors, {\em ASIACRYPT (1)}, volume
  8269 of {\em Lecture Notes in Computer Science}, pages 117--133. Springer,
  2013.

\bibitem{milnor2}
John Milnor.
\newblock Symmetric inner products in characteristic 2.
\newblock {\em Prospects in Mathematics”, Annals of Math. Studies},
  70:59--75, 1971.

\bibitem{milnorhusemoller}
John~W. Milnor and Dale Husemoller.
\newblock {\em {Symmetric bilinear forms}}.
\newblock Springer, Berlin, Heidelberg, Paris, 1973.

\bibitem{neukirch1999algebraic}
J{\"u}rgen Neukirch.
\newblock {\em {Algebraic number theory}}.
\newblock Springer, 1999.

\bibitem{omeara}
Timothy O'Meara.
\newblock {\em {Introduction to quadratic forms}}.
\newblock Classics in mathematics. Springer, Berlin, Heidelberg, Paris, 2000.
\newblock Reprint of the 1973 edition.

\bibitem{DBLP:conf/eurocrypt/Patarin96}
Jacques Patarin.
\newblock {Hidden Fields Equations (HFE) and Isomorphisms of Polynomials (IP):
  Two New Families of Asymmetric Algorithms}.
\newblock In Ueli~M. Maurer, editor, {\em EUROCRYPT}, volume 1070 of {\em
  Lecture Notes in Computer Science}, pages 33--48. Springer, 1996.

\bibitem{DBLP:conf/eurocrypt/PatarinGC98}
Jacques Patarin, Louis Goubin, and Nicolas Courtois.
\newblock Improved algorithms for isomorphisms of polynomials.
\newblock In Kaisa Nyberg, editor, {\em EUROCRYPT}, volume 1403 of {\em Lecture
  Notes in Computer Science}, pages 184--200. Springer, 1998.

\bibitem{DBLP:conf/eurocrypt/Perret05}
Ludovic Perret.
\newblock {A Fast Cryptanalysis of the Isomorphism of Polynomials with One
  Secret Problem}.
\newblock In Ronald Cramer, editor, {\em EUROCRYPT}, volume 3494 of {\em
  Lecture Notes in Computer Science}, pages 354--370. Springer, 2005.

\end{thebibliography}
\ifapx \appendix
\section{An algorithm reducing a matrix pencil to Kronecker normal form}%<<<1
\label{A:algo}

We explain how, given a pencil~$B = (B_{0}, B_{∞})$ in matrix form, we
can compute the Kronecker blocks of~$B$. This algorithm is a direct
translation of part~\ref{SS:Kronecker-reduction}.
\newcounter{step}
\def\step#1{\paragraph{Step \arabic{step}: #1.}\stepcounter{step}}

\step{Compute the minimal isotropic vectors~$(e_1,…,e_h)$}%<<<2

Write the~$n × (2n)$-matrix~$(B_{∞} \; B_0)$ in lower row echelon form as
\begin{equation}
(B_{∞} \; B_0) \;=\; (⋆) · \mat{A_{∞} & 0\\ C & A_0},
\end{equation}
where $(⋆)$~is an invertible~$n × n$-matrix and
$0$~is a $r × n$-block with $r$~being the largest possible value.
This implies that the $n-r$~lines of~$A_0$ are linearly independent, and
therefore that its columns have full rank~$n-r$; therefore, there exists
a matrix~$F$ such that~$C = -A_0 F$. From this, we see that
\begin{equation}
B_{∞} x + B_0 y = 0 \;⇔ \; \begin{cases} A_{∞} x = 0\\ y ∈ Fx + \Ker A_0.
\end{cases}
\end{equation}
In particular, the case~$x = 0$ tells us that $\Ker A_0 = \Ker B_0$. By
using the \emph{upper} row echelon form, we likewise compute a matrix~$G$
such that~$B_{∞} x = B_0 y$ implies~$x ∈ G y + \Ker B_{∞}$.

A \emph{chain} of length~$h$ is a solution~$(e_0,…,e_h)$ of the
equations~$b_0(e_i) + b_{∞}(e_{i-1}) = 0$ and~$b_{∞}(e_h) = 0$.
We define by induction a sequence~$(U_i)$ of vector spaces such that the
$h$-chains are defined by the relations~$e_0 ∈ U_h$ and~$e_i ∈ F(e_{i-1})
+ U_{h-i}$.

The base case is that of chains of length~$0$, which are the
elements of~$\Ker B_{∞}$. We define $U_0 = \Ker B_{∞}$.

A $(h+1)$-chain is a $(h+2)$-uple~$(e_0,…,e_{h+1})$ such that
$(e_1,…,e_{h+1})$~is a $h$-chain and $b_{0} e_1 + b_{∞} e_0 = 0$.
The first condition amounts to~$e_1 ∈ U_h$ and~$e_i = f(e_{i-1}) +
U_{h-i}$ for all~$i ≥ 2$; the second one means that $e_0 ∈ G U_{h} + \Ker
B_{∞}$. We define~$U_{h+1} = G U_h + \Ker B_{∞}$.

This allows us to compute minimal isotropic vectors of length~$h$ as
satisfying the relations
\begin{equation}\label{eq:m.i.v.-matrix}
e_0 ∈ U_{h} ∩ \Ker B_0, \quad e_{i} ∈ F e_{i-1} + U_{h-i}.
\end{equation}
This determines the space of isotropic vectors of degree~$≤ h$ with
total complexity~$O(n^3 (h+1))$; moreover, as this computation is
triangular, it also gives the space of isotropic vectors of degree~$≤ h'$
for all~$h' ≤ h$, so that we only need to perform one run of this
algorithm over all the singular part of the pencil. Once we have isolated
the minimal Kronecker module~$K$ of~$V$ in Step 2, we may then project
this basis on the quotient~$V/K$ to directly obtain a (sorted) basis of
the isotropic vectors of~$V/K$.

\step{Compute a Kronecker module as a direct factor}%<<<2
Given the vectors~$e_1,…,e_h$ computed in the previous step,
we can, with no more than~$O(n^3)$ field operations, compute
vectors~$f_1,…,f_h$ such that~$b_0(e_i, f_j) = 1$ if~$i = j$ and
$0$~otherwise.
In any basis completing the family~$(e_0, …, e_h; f_1, …, f_h)$,
the symmetric pencil~$(b_{λ})$ has the matrix
\begin{equation}\label{eq:matrix-b1}
B_{λ} = \mat{0 & K'_{λ} & 0\\\transpose{K'_{λ}} & A_{λ}&\transpose{C_{λ}}\\
  0 & C_{λ} & B'_{λ}},
\end{equation}
where the blocks have size~$d+1$, $d$ and~$n-(2d+1)$. We use a change of
coordinates of the form
\begin{equation}\label{eq:p1}
P = \mat{1 & 0 & X\\0&1&0\\0&Y&1}.
\end{equation}
The action of~$P$ on the sub-matrix~$C_{λ}$ of~$B_{λ}$ is given by $C_{λ}
← C_{λ} + \transpose{X} K'_{λ} + B_{λ} Y$. Now let~$x_0,…,x_{h}$;
$y_1,…,y_h$; $c_1,…,c_h$; $c'_1,…,c'_h$ be the columns of~$\transpose{X}, Y,
C_0$ and~$C_{∞}$. We then have to
solve the equations
\begin{equation}\label{eq:chvar-x}
\begin{cases} c'_i + x_i + B'_0 y_i = 0\\ i = 1,…,h\end{cases};\quad
\begin{cases} c'_i + x_{i-1} + B'_∞ y_i = 0\\i=1,…,h\end{cases}.
\end{equation}
This uniquely determines the values~$x_0$ and~$x_{h}$. The equations
for~$x_1,…,x_{h-1}$ have solutions iff the values~$y_1,…,y_h$ satisfy
the relations
$B'_0 y_i + B'_{∞} y_{i+1} = c'_{i+1} - c_i$ for~$i = 1,…,h-1$. This
translates into matrix form as
\begin{equation}\label{eq:chvar-y}
\mat{B'_0&B'_{∞}&&0\\&\;\sddots{B'_{∞}}&\sddots{B'_{∞}}&\\0&&B'_0&B'_{∞}} ·
  \mat{y_1\\⋮\\y_h} = \mat{c'_2-c_1\\⋮\\c'_h-c_{h-1}}.
\end{equation}
We may solve this equation using the same technique as that of Step~1,
for a total cost of~$O(n^3 h)$. Computing the values~$(x_i)$ is then
straightforward.

\step{Put the Kronecker module in canonical form}%<<<2
We now compute a coordinate change
\begin{equation}\label{eq:chvar-q}
\def\arraystretch{.7}
Q = \mat{1&Z&\\&1\\&&1}
\end{equation}
such that~$\transpose{Q} · B_{λ} · Q$~puts the Kronecker module in the
canonical form~$K_{h}$ described in~\eqref{eq:def-K}. The action of~$Q$
on~$A_{λ}$ is given by~$A_{λ} ← A_{λ} + \transpose{Z} K_{λ} +
\transpose{K_{λ}} Z$. Let~$Z = (z_{i,j})$, $A_0 = (a_{i,j})$ and~$A_{∞} =
a'_{i,j}$; the equations to solve are then
\begin{equation}\label{eq:zij}
z_{i,j} + z_{j,i} = a_{i,j}, \quad z_{i-1,j} + z_{j-1,i} = a'_{i,j},
\quad \text{for~$i,j = 1,…,h$}.
\end{equation}
Since the matrices~$A_{0}$ and~$A_{∞}$ are symmetric, only the
equations for~$i ≥ j$ are relevant. We derive from~$z_{i,j}$ the
values $z_{i,i} = \frac{1}{2} a_{i,i}$ and~$z_{i-1,i} = \frac{1}{2}
a'_{i,i}$. The remaining values~$z_{i,j}$ for~$i+j=\mathrm{constant}$ are
deduced from the relation~$z_{i,j} - z_{i-1,j+1} = a_{i,j} - a'_{i,j+1}$.
We check that these relations compute all the values~$z_{i,j}$ in optimal
time, which is~$O(n^2)$ computations in the base field~$K$.

\fi

\end{document}